\documentclass[aps,10pt,pra,twocolumn,showpacs,nofootinbib,superscriptaddress]{revtex4-1}
\usepackage{amsfonts}
\usepackage{amsmath}
\usepackage{amssymb}
\usepackage{graphicx}

\usepackage{pict2e}
\usepackage{diagbox}

\usepackage{color}
\usepackage{cancel}
\usepackage{ulem}

\def\filetype{eps}

\begin{document}
\title{Hartree-Fock treatment of Fermi polarons using the Lee-Low-Pine transformation}
\author{Ben Kain}
\affiliation{Department of Physics, College of the Holy Cross, Worcester, Massachusetts 01610 USA}
\author{Hong Y.\ Ling}
\affiliation{Department of Physics and Astronomy, Rowan University, Glassboro, New Jersey 08028 USA}


\begin{abstract}
\noindent We consider the Fermi polaron problem at zero temperature, where a single impurity interacts with non-interacting host fermions.  We approach the problem starting with a Fr\"{o}hlich-like Hamiltonian where the impurity is described with canonical position and momentum operators.  We apply the Lee-Low-Pine (LLP) transformation to change the fermionic Fr\"{o}hlich Hamiltonian into the fermionic LLP Hamiltonian which describes a many-body system containing host fermions only.  We adapt the self-consistent Hartree-Fock (HF) approach, first proposed by Edwards, to the fermionic LLP Hamiltonian in which a pair of host fermions with momenta $\mathbf{k}$ and $\mathbf{k}^{\prime}$ interact with a potential proportional to $\mathbf{k}\cdot\mathbf{k}^{\prime}$.  We apply the HF theory, which has the advantage of not restricting the number of particle-hole pairs, to repulsive Fermi polarons in one dimension.  When the impurity and host fermion masses are equal our variational ansatz, where HF orbitals are expanded in terms of free-particle states, produces results in excellent agreement with McGuire's exact analytical results based on the Bethe ansatz. This work raises the prospect of using the HF ansatz and its time-dependent generalization as building blocks for developing all-coupling theories for both equilibrium and nonequilibrium Fermi polarons in higher dimensions. 
\end{abstract}
\maketitle


\section{Introduction}

A polaron is an impurity in a host medium, dressed with excitations of the host medium particles with which it interacts.  A polaron may be classified as a Bose polaron or a Fermi polaron depending on whether the host particle excitations obey Bose or Fermi statistics. The idea traces its roots to more than half a century ago when Landau and Pekar
\cite{landau46ZhEkspTeorFiz.16.341,landau48ZhEkspTeorFiz.18.419} likened a conduction electron dressed with phonons (bosons) in an ionic crystal to a  polaron.  In condensed matter physics, then, polaron studies began with Bose polarons and later spread to impurities submerged in a bath of fermions, e.g.\ Anderson's orthogonality catastrophe
\cite{anderson67PhysRevLett.18.1049,mahan00ManyParticlePhysicsBook}, the Kondo
effect
\cite{kondo64ProgressOfTheoreticalPhysics.32.37,hewson97HeavyFermionsBook},
and the motion of ions in liquid $^{3}$He
\cite{kondo83JournalOfLowTemperaturePhysic.50.21}.

The advent of cold atom systems, with their unprecedented flexibility, has greatly heightened the prospect that polaron properties may be explored to great precision across a broad interaction regime and with different dimensionality.  The recent renaissance in the study of polarons began with Fermi polarons (see
\cite{chevy10RepProgPhys.73.112401,massignan14ReportsOnProgressInPhysics.77.034401}
for a review), as cold atom polarons were first realized in experiments with
mixtures of highly imbalanced Fermi gases
\cite{schirotzek09PhysRevLett.102.230402,nascimbene09PhysRevLett.103.170402}.  The results of these experiments were found to be in excellent agreement with earlier theoretical predictions
\cite{chevy06PhysRevA.74.063628,lobo06PhysRevLett.97.200403,combescot07PhysRevLett.98.180402,combescot08PhysRevLett.101.050404,prokofev08PhysRevB.77.020408}, which were inspired by experimental realizations of imbalanced mixtures of cold fermionic atoms
\cite{zwierlein06Science.311.492,partridge06Science.31.503}. This same period also witnessed an increased interest in Bose polarons where the role of the host medium is played by a Bose-Einstein condensate (BEC)
\cite{astrakharchik04PhysRevA.70.013608,cucchietti06PhysRevLett.96.210401,kalas06PhysRevA.73.043608,bruderer08EurPhysLett.82.30004,pilati08PhysRevLett.100.030401,Huang09ChinesePhysicsLetters.26.080302,tempere09PhysRevB.80.184504}.
In more recent years, there has been a plethora of activity aimed at
understanding Fermi polarons both theoretically
\cite{zwerger09PhysRevA.80.053605,cui10PhysRevA.81.041602,pilati10PhysRevLett.105.030405,massignan11EurPhysJD.65.83,schmidt11PhysRevA.83.063620,
Mathy,
guan12FrontierOfPhysics.7.8,yi15PhysRevA.92.013620,Mao16PhysRevA.94.043645,schmidt17arXiv:1702.08587}
and experimentally
\cite{marco12Nature.485.619,zhang12PhysRevLett.108.235302,kohstall12Nature.485.615,cetina16Science.354.6308,scazza17PhysRevLett.118.083602}
and Bose polarons both theoretically \cite{casteels11LaserPhysics.21.1480,casteels12PhysRevA.86.043614,rath13PhysRevA.88.053632,kain14PhysRevA.89.023612,
shashi14PhysRevA.89.053617,li14PhysRevA.90.013618,grusdt15ScientificReports.5.12124,ardila15PhysRevA.92.033612,
vlietinck15NewJournalOfPhysics.17.033023,shchadilova16PhysRevA.93.043606,sogaard15PhysRevLett.115.160401,
levinsen15PhysRevLett.115.125302,kain16PhysRevA.94.013621,schmidt16PhysRevLett.116.105302,shchadilova16PhysRevLett.117.113002,
Grusdt17arXiv:1704.02605,Grusdt17arXiv:1704.02606,  Sun, Sun2} and experimentally
\cite{catani12PhysRevA.85.023623,hu16PhysRevLett.117.055301,jorgensen16PhysRevLett.117.055302}.  As such, polarons continue to be a topic of intense current interest to the cold atom physics community.

Cross-field fertilization has been a hallmark of physics research. The study of Bose polarons owes its rapid development, in part, to the remarkable progress made in developing and improving our understanding of the
Fermi polaron problem in the past decade.  For instance, a Chevy-like 
variational ansatz \cite{chevy06PhysRevA.74.063628} originally
developed for the Fermi polaron problem has been adapted
successfully to its bosonic counterpart
\cite{li14PhysRevA.90.013618,levinsen15PhysRevLett.115.125302}.  In view of
recent progress made in the study of Bose polarons, we investigate the
opposite scenario where the study of Fermi polarons takes cues from its bosonic cousin. 

Specifically we look to the Fr\"{o}hlich Hamiltonian, popular in the study of
Bose polarons, where the impurity is treated, from the outset, as a
single-particle quantum system described with canonical position and
momentum operators.  This may be contrasted with the usual Hamiltonian employed in the study of Fermi polarons, where both impurities and host fermions are described in the language of second quantization for many-body quantum systems. An advantage to using the Fr\"{o}hlich Hamiltonian is that the Lee-Low-Pine (LLP) transformation \cite{lee53PhysRev.90.297} can be easily applied, which eliminates the impurity degree of freedom yielding a Hamiltonian describing a single component system with host particles only.

The LLP transformation amounts to changing from the laboratory frame to a moving frame and is therefore a general transformation not limited to Fr\"{o}hlich Bose polarons. In fact, Edwards \cite{Edwards1990PThPS.101.453}
recognized years ago for Fermi polarons that moving to a frame attached to the impurity has distinct advantages.  Such a change of frame has since found applications in analogous problems across different contexts (see, for example,
\cite{Castella93PhysRevB.47.16186,Lamacraft09PhysRevB.79.241105,Mathy12NaturePhysics.8.881,edward13JournalOfPhysics.25.425602,grusdt17NatCommun.7.11994}).  In both the Bose and the Fermi Fr\"{o}hlich models, the host particles after the LLP transformation are found to interact with a two-body interaction quite different from the usual two-body interaction.  This induced interaction, which is absent in the lab frame, has been the focus of much recent research in Bose polarons
\cite{grusdt15ScientificReports.5.12124,shchadilova16PhysRevA.93.043606,kain16PhysRevA.94.013621}, which gives strong support that the LLP transformation in combination with many-body quantum field theory constitutes a powerful tool for developing all-coupling theories for Fr\"{o}hlich polarons.

In this work, motivated by these latest developments in the study of Bose polarons, we formulate the Fermi polaron problem in a language familiar to the study of Fr\"{o}hlich polarons involving the use of the LLP transformation with the goal of developing an all coupling theory for Fermi polarons.

In Sec.\ II, we introduce the fermionic analog of the Fr\"{o}hlich Hamiltonian, which describes a single spin-$\downarrow$
impurity interacting with \textit{non-interacting} spin-$\uparrow$ fermions, and then apply the LLP transformation to obtain
the fermionic LLP model, which describes a spin polarized Fermi system
containing \textit{interacting} spin-$\uparrow$ fermions (and free of the impurity degrees of freedom).

In Sec.\ III, we show that the fermionic LLP model naturally embraces Chevy's ansatz
\cite{chevy06PhysRevA.74.063628}, which is a superposition of many-body states with
various numbers of particle-hole excitations, as a variational ansatz for weak coupling Fermi polarons.

In Sec.\ IV, we adapt the Hartree-Fock (HF) variational ansatz, where fermions at equilibrium are assumed to be in a Slater determinant, to the fermionic LLP model of arbitrary dimension. The (HF) ansatz has the advantage that it does not limit the number of particle-hole pairs and is thus expected to be more accurate, particularly in the strong coupling regime, than its perturbative
analog presented in Sec.\ III where the number of particle-hole pairs is fixed a priori.

In Sec.\ V, we establish the validity of our approach by applying the general theory in Sec.\ IV to a repulsive Fermi polaron in a quasi-one-dimensional (quasi-1D) setting.  For the case where the impurity and host fermion masses are equal, the polaron energy and
effective mass are found to be in excellent agreement with McGuire's analytical results \cite{McGuire65JMathPhys.6.432,McGuire66JMathPhys.7.123} based on the Bethe ansatz across a wide range of interaction strengths.  We perform an in-depth analysis of the single-particle spectrum in the moving frame, revealing novel features that distinguish weakly-coupled Fermi polarons from strongly-coupled ones.   We also discuss and benchmark results where the impurity mass is different from the host fermion mass. 

In Sec.\ VI, we summarize our results and provide further comments about our approach to the Fermi polaron problem.


\section{Fermi Polaron Hamiltonian after Lee-Low-Pine Transformation}

We start with the Hamiltonian for a two-component Fermi
gas mixture in the limit of short-range interactions:
\begin{equation} \hat{H}^{\prime}=\sum_{\mathbf{k}}\left(  \epsilon_{\mathbf{k}}\hat
{a}_{\mathbf{k}}^{\dag}\hat{a}_{\mathbf{k}}+\epsilon_{\mathbf{k}}^{I}\hat
{c}_{\mathbf{k}}^{\dag}\hat{c}_{\mathbf{k}}\right)  +\frac{g}{\mathcal{V}}%
\sum_{\mathbf{kk}^{\prime}\mathbf{q}}\hat{c}_{\mathbf{k}+\mathbf{q}}^{\dag
}\hat{a}_{\mathbf{k}^{\prime}-\mathbf{q}}^{\dag}\hat{a}_{\mathbf{k}^{\prime}
}\hat{c}_{\mathbf{k}}, \label{H prime}
\end{equation}
where $\mathcal{V}$ is the quantization ``volume," $g$ is the $s$-wave interaction strength between an impurity atom and a host fermion, and $\hat{a}_{\mathbf{k}}$ and $\hat{c}_{\mathbf{k}}$ denote, respectively,
annihilation operators for a spin-$\uparrow$ majority (host) fermion of mass
$m$ with kinetic energy $\epsilon_{\mathbf{k}}=k^{2}/2m$ and a
spin-$\downarrow$ minority (impurity) atom of mass $m_{I}$ with kinetic
energy $\epsilon_{\mathbf{k}}^{I}=k^{2}/2m_{I}$, where $\mathbf{k}$ is the momentum.  Taking the so-called single impurity limit, we 
make the replacements $\sum_{\mathbf{k}}\hat{c}_{\mathbf{k}+\mathbf{q}
}^{\dag}\hat{c}_{\mathbf{k}}\rightarrow\exp(  i\mathbf{q}\cdot
\mathbf{\hat{r}})  $ and $\sum_{\mathbf{k}}\epsilon_{\mathbf{k}}^{I}%
\hat{c}_{\mathbf{k}}^{\dag}\hat{c}_{\mathbf{k}}\rightarrow\hat{p}^{2}/(2m_{I})$, which eliminates the impurity field operators in favor of the impurity position and momentum operators,
$\mathbf{\hat{r}}$ and $\mathbf{\hat{p}}$, transforming Eq.\
(\ref{H prime}) into
\begin{equation}
\hat{H}^{\prime}=\frac{\hat{p}^{2}}{2m_{I}}+\sum_{\mathbf{k}}\epsilon
_{\mathbf{k}}\hat{a}_{\mathbf{k}}^{\dag}\hat{a}_{\mathbf{k}}+\frac
{g}{\mathcal{V}}\sum_{\mathbf{k},\mathbf{q}}e^{i\mathbf{q\cdot \hat{r}}}\hat
{a}_{\mathbf{k}-\mathbf{q}}^{\dag}\hat{a}_{\mathbf{k}},
\label{Fermi Frohlich Hamiltonian}%
\end{equation}
which is the fermionic analog of the well-known Fr\"{o}hlich Hamiltonian.

The LLP transformation is based upon total momentum conservation.  For our Fermi model, a simple evaluation finds $[
\mathbf{\hat{p}},\hat{H}^{\prime}]  =-[  \mathbf{\hat{p}}_{f}
,\hat{H}^{\prime}]  \neq0$, indicating that the
impurity momentum $\mathbf{\hat{p}}$ and the total fermion momentum
\begin{equation}
\mathbf{\hat{p}}_{f}=\sum_{\mathbf{k}}\mathbf{k}\hat{a}_{\mathbf{k}}^{\dag
}\hat{a}_{\mathbf{k}}
\end{equation}
are not conserved separately but that their sum is conserved,
$[ \mathbf{\hat{p}}+\mathbf{\hat{p}}_{f},\hat{H}^{\prime}]  =0$.  This is to be expected since the impurity together with the background fermions forms an isolated system.  Just as in the Bose polaron problem, we now introduce the fermionic LLP transformation:
\begin{equation}
\hat{S}=e^{i\mathbf{\hat{r}\cdot}\sum_{\mathbf{k}}\mathbf{k}\hat
{a}_{\mathbf{k}}^{\dag}\hat{a}_{\mathbf{k}}}. \label{LLP transformation}
\end{equation}

A comment regarding the effect of $\hat{S}$ on vacuum states is in order.  For Bose polarons at zero temperature, the vacuum  is empty (i.e.\ free of phonons) and is invariant under the LLP transformation.  By contrast, for our Fermi system at zero temperature the ``vacuum" is not empty and corresponds to a filled Fermi sea where states below the Fermi energy $\epsilon_{F} =k_{F}^{2}/2m$ (or Fermi momentum $k_F$) are all occupied.  Nevertheless, because the Fermi sea is inert, i.e\ the total fermion momentum of the Fermi sea is zero, the Fermi vacuum is also invariant under the LLP transformation.  (Unless otherwise stated, by ``Fermi sea" we always mean a non-interacting Fermi sea.)

It is easily verified that $\mathbf{\hat{p}}$ and $\hat{a}_{\mathbf{k}}$ transform under $\hat{S}$ analogously to the Bose polaron problem,
\begin{align}
\hat{S}^{-1}\mathbf{\hat{p}}\hat{S}  &  =\mathbf{\hat{p}}-\sum_{\mathbf{k}%
}\mathbf{k}\hat{a}_{\mathbf{k}}^{\dag}\hat{a}_{\mathbf{k}},\\
\hat{S}^{-1}\hat{a}_{\mathbf{k}}\hat{S}  &  =\hat{a}_{\mathbf{k}%
}e^{-i\mathbf{k\cdot\hat{r}}}.
\end{align}
The Hamiltonian (\ref{Fermi Frohlich Hamiltonian}) under the LLP transformation, $\hat{H}=\hat{S}^{-1}\hat
{H}^{\prime}\hat{S},$ then reads
\begin{equation}
\hat{H}=\frac{\left(  \mathbf{p}-\sum_{\mathbf{k}}\mathbf{k}\hat
{a}_{\mathbf{k}}^{\dag}\hat{a}_{\mathbf{k}}\right)  ^{2}}{2m_{I}}%
+\sum_{\mathbf{k}}\epsilon_{\mathbf{k}}\hat{a}_{\mathbf{k}}^{\dag}\hat
{a}_{\mathbf{k}}+\frac{g}{\mathcal{V}}\sum_{\mathbf{k},\mathbf{k}^{\prime}%
}\hat{a}_{\mathbf{k}}^{\dag}\hat{a}_{\mathbf{k}^{\prime}}, \label{LLP H}%
\end{equation}
which we refer to as the fermionic LLP Hamiltonian.

The LLP transformation (\ref{LLP transformation}) is a Galilean boost operator and can thus be regarded qualitatively as boosting the system into a frame moving at a speed determined by the total fermion momentum.  In this frame the impurity momentum $\mathbf{\hat{p}}$ is a constant of motion allowing it to be replaced with the $c$-number $\mathbf{p}$ in Eq.\ (\ref{LLP H}).  Comparing the Hamiltonian in Eq.\ (\ref{LLP H}) with the Hamiltonian prior to the LLP transformation in Eq.\ (\ref{Fermi Frohlich Hamiltonian}), we see that the post-LLP Hamiltonian describes a system containing only fermions, but at the expense of introducing an interaction between them.  This is entirely analogous to the LLP transformation in the Bose polaron problem.


\section{Perturbative variational ansatz: Chevy's ansatz}

Polaron problems, be they bosonic or fermionic, are usually studied with variational methods.  As a matter of fact, the LLP transformation was originally introduced as a first step towards developing a variational approach to weakly-coupled Fr\"{o}hlich polarons \cite{lee53PhysRev.90.297}.  To illustrate how the fermionic LLP transformation may inspire variational approaches to weak coupling Fermi polarons, we divide momentum states into particle states
$\mathbf{k}$ with $|\mathbf{k}| >k_{F}$ and hole states
$\mathbf{q}$ with $|\mathbf{q}| <k_{F}$ and introduce the
canonical particle-hole transformation,
\begin{equation}
\hat{b}_{\mathbf{k}}\equiv\hat{a}_{\mathbf{k}}, \quad
\hat{b}_{\mathbf{k}}^{\dag
}\equiv\hat{a}_{\mathbf{k}}^{\dag},
\quad \hat{b}_{\mathbf{q}}\equiv\hat
{a}_{\mathbf{q}}^{\dag},
\quad \hat{b}_{\mathbf{q}}^{\dag}\equiv\hat{a}_{\mathbf{q}
}, \label{particle-hole}
\end{equation}
where $\hat{b}_{\mathbf{k}}$ ($\hat{b}_{\mathbf{k}}^{\dag}$) annihilates (creates) a particle with momentum $\mathbf{k}$
and energy $\epsilon_{\mathbf{k}}$ and $\hat{b}_{\mathbf{q}}$ $(
\hat{b}_{\mathbf{q}}^{\dag})$  annihilates (creates) a hole with momentum $\mathbf{-q}$ and energy $-\epsilon
_{\mathbf{q}}$.  The application of Eq.\ (\ref{particle-hole}) changes the LLP Hamiltonian (\ref{LLP H}) into
\begin{equation}
\hat{H}=\hat{H}_{0}+\hat{V},
\end{equation}
where
\begin{align}
\hat{H}_{0}  & =\frac{\left[  \mathbf{p}-\sum_{\mathbf{k}}\mathbf{k}\hat
{b}_{\mathbf{k}}^{\dag}\hat{b}_{\mathbf{k}}-\sum_{\mathbf{q}}\left(
-\mathbf{q}\right)  \hat{b}_{\mathbf{q}}^{\dag}\hat{b}_{\mathbf{q}}\right]
^{2}}{2m_{I}}\nonumber\\
&\qquad+ \sum_{\mathbf{q}}\epsilon_{\mathbf{q}}+\sum_{\mathbf{k}}\epsilon
_{\mathbf{k}}\hat{b}_{\mathbf{k}}^{\dag}\hat{b}_{\mathbf{k}}+\sum_{\mathbf{q}%
}\left(  -\epsilon_{\mathbf{q}}\right)  \hat{b}_{\mathbf{q}}^{\dag}\hat
{b}_{\mathbf{q}} \label{H_0}%
\end{align}
follows from the first two terms in Eq. (\ref{LLP H}) with the final term giving
\begin{align}
\hat{V}  & = gn_{F} \label{H_S}\\ \notag
&+\frac{g}{\mathcal{V}}  \left(  \sum_{\mathbf{k},\mathbf{k}^{\prime}}\hat{b}_{\mathbf{k}}^{\dag
}\hat{b}_{\mathbf{k}^{\prime}}-\sum_{\mathbf{q},\mathbf{q}^{\prime}}
\hat
{b}_{\mathbf{q}}^{\dag}\hat{b}_{\mathbf{q}^{\prime}}+\sum_{\mathbf{k}%
,\mathbf{q}}\hat{b}_{\mathbf{k}}^{\dag}\hat{b}_{\mathbf{q}}^{\dag}%
-\sum_{\mathbf{k},\mathbf{q}}\hat{b}_{\mathbf{k}}\hat{b}_{\mathbf{q}}\right),
\end{align}
where $n_{F}$ is the fermion number density.  Here, $\hat{H}_{0}$ in Eq.\ (\ref{H_0}) represents the sum of the kinetic energy associated with the impurity recoil (first line) and the free fermion energy (second line).  A particularly attractive feature of the LLP Hamiltonian (\ref{LLP H}) is that the $s$-wave interaction between an impurity and a host fermion manifests simply as a localized
impurity potential, $g\delta(\mathbf{r})$, embedded in host fermions.  $\hat{V}$ in Eq.\ (\ref{H_S}) represents this potential scattering.

\begin{figure}
\centering
\includegraphics[width=3.25in]{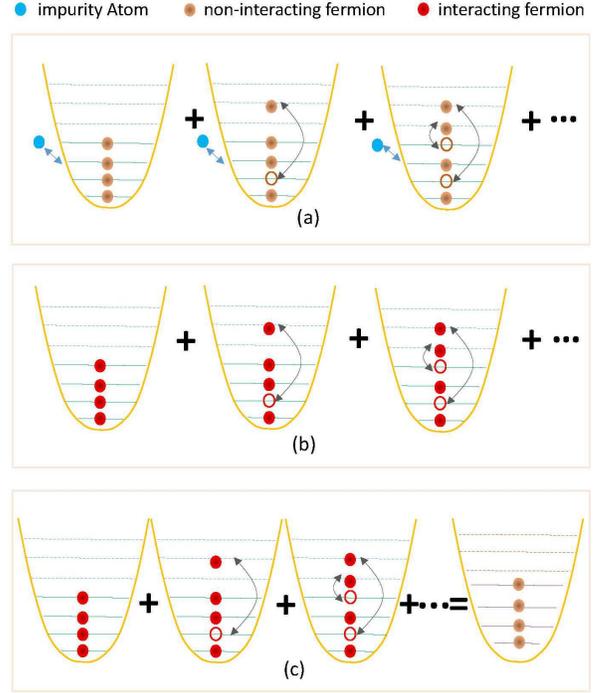}
\caption{(a) Chevy's variational ansatz based on the standard second-quantized Hamiltonian (\ref{H prime}) in the lab frame.  (b) Chevy's variational ansatz based on the LLP Hamiltonian (\ref{LLP H}) in the moving frame.  (c) Self-consistent Hartree-Fock variational ansatz.}
\label{Fig: variationalAnsatz}
\end{figure}

The eigenstates of $\hat{H}_{0}$ are particle-hole excitations.  Not only does this fact facilitate the implementation of a perturbation theory in which $\hat{H}_0$ in Eq.\ (\ref{H_0}) and $\hat{V}$ in Eq.\ (\ref{H_S}) are treated as the unperturbed Hamiltonian and its perturbation, but it also motivates a variational approach in which a trial wave function is a superposition of families of states grouped according to the number of particle-hole pairs as illustrated in Fig.\
\ref{Fig: variationalAnsatz}(b).  Each term in the $i$th family consists of $i$ particle-hole pairs. For example, the trial wave function up to two particle-hole pairs is
\begin{align}
\left\vert \psi\right\rangle  
&=\alpha_{0}\left\vert 0\right\rangle
+\sum_{\mathbf{kq}}\alpha_{\mathbf{kq}}\left\vert 1_{\mathbf{k}}1_{\mathbf{q}%
}\right\rangle \nonumber\\
&\qquad +  \frac{1}{\left(  2!\right)  }\sum_{\mathbf{kk}^{\prime}\mathbf{qq}^{\prime
}}\alpha_{\mathbf{kk}^{\prime}\mathbf{qq}^{\prime}}\left\vert 1_{\mathbf{k}%
}1_{\mathbf{k}^{\prime}}1_{\mathbf{q}}1_{\mathbf{q}^{\prime}}\right\rangle ,
\label{trial wave 1}
\end{align}
where $\alpha_{0}$, $\alpha_{\mathbf{kq}}$, and
$\alpha_{\mathbf{kk}^{\prime}\mathbf{qq}^{\prime}}$ are variational
parameters and $| 1_{\mathbf{k}}1_{\mathbf{q}}\rangle
\equiv\hat{a}_{\mathbf{k}}^{\dag}\hat{a}_{\mathbf{q}}| 0\rangle
$ and $| 1_{\mathbf{k}}1_{\mathbf{k}^{\prime}}1_{\mathbf{q}%
}1_{\mathbf{q}^{\prime}}\rangle \equiv\hat{a}_{\mathbf{k}}^{\dag}\hat
{a}_{\mathbf{k}^{\prime}}^{\dag}\hat{a}_{\mathbf{q}}\hat{a}_{\mathbf{q}
^{\prime}}| 0\rangle $ are Slater
determinants for one and two particle-hole excitations on top of the
Fermi sea $| 0\rangle $.

The same variational expansion for Hamiltonian (\ref{H prime})
takes the form
\begin{align}
\left\vert \psi^{\prime}\right\rangle   
&=\alpha_{0}\hat{c}_{\mathbf{p}%
}^{\dag}\left\vert 0\right\rangle +\sum_{\mathbf{kq}}\alpha_{\mathbf{kq}}%
\hat{c}_{\mathbf{p}+\mathbf{q}-\mathbf{k}}^{\dag}\left\vert 1_{\mathbf{k}%
}1_{\mathbf{q}}\right\rangle \nonumber\\
&\quad +  \frac{1}{\left(  2!\right)  }\sum_{\mathbf{kk}^{\prime}\mathbf{qq}^{\prime
}}\alpha_{\mathbf{kk}^{\prime}\mathbf{qq}^{\prime}}\hat{c}_{\mathbf{p}%
+\mathbf{q+q}^{\prime}-\mathbf{k}-\mathbf{k}^{\prime}}^{\dag}\left\vert
1_{\mathbf{k}}1_{\mathbf{k}^{\prime}}1_{\mathbf{q}}1_{\mathbf{q}^{\prime}%
}\right\rangle , \label{|psi'>}%
\end{align}
which is Chevy's celebrated ansatz first introduced by Chevy for one
particle-hole pair \cite{chevy06PhysRevA.74.063628} and subsequently
generalized to higher particle-hole pairs
\cite{combescot08PhysRevLett.101.050404}. Figure \ref{Fig:
variationalAnsatz}(a) illustrates the expansion used to construct the trial wave function (\ref{|psi'>}).

Because particle-hole excitations, such as $\left\vert
1_{\mathbf{k}}1_{\mathbf{q}}\right\rangle $, are eigenstates of the total
fermion momentum, one can show that Chevy's ansatz (\ref{|psi'>}) transforms under the LLP transformation as 
\begin{equation}
\hat{S}\left\vert \psi^{\prime}\right\rangle =\left\vert \mathbf{p}%
\right\rangle \otimes\left\vert \psi\right\rangle , \label{psi'-psi}%
\end{equation}
which is the direct product of $| \mathbf{p}\rangle $, an
impurity state with total momentum $\mathbf{p}$, and $|
\psi\rangle $ in Eq.\ (\ref{trial wave 1}), our trial wave function for host
fermions only, demonstrating again that the LLP transformation decouples the impurity from the host fermions.

The necessity to terminate the expansion as in our example in Eq.\ (\ref{trial wave 1}) can be traced to $\hat{V}$ in Eq.\ (\ref{H_S}).  The term in the top line shifts the mean-field energy.  In the second line the first term couples particle states and the second term couples hole states, both of which couple within the \textit{same} particle-hole family.  The final two terms on the last line couple \textit{different} families through particle-hole creation (third term) and particle-hole annihilation (fourth term) and it is these couplings (across different families) that make it necessary to restrict the number of particle-hole pairs in the trial wave function so that the problem can be described by a closed set of equations. 


\section{Non-perturbative all-coupling variational ansatz: Self-consistent Hartree-Fock ansatz}

A Fermi polaron is simply an impurity clothed with
particle-hole excitations.  In order to yield a more accurate polaron description, efforts were made early on to construct nonperturbative variational ansatzes which do not restrict the number of particle-hole
excitations
\cite{combescot07PhysRevLett.98.180402,combescot08PhysRevLett.101.050404}.  Here, we aim to achieve this goal by using an approach inspired in large part by recent advancements in the study of Bose polarons
\cite{grusdt15ScientificReports.5.12124,shchadilova16PhysRevA.93.043606,kain16PhysRevA.94.013621}.  We begin by casting Eq.\ (\ref{LLP H}) into the normally-ordered form
\begin{align}
\hat{H}  &  =\frac{p^{2}}{2m_{I}}+\frac{g}{\mathcal{V}}\sum_{\mathbf{k}%
,\mathbf{k}^{\prime}}\hat{a}_{\mathbf{k}}^{\dag}\hat{a}_{\mathbf{k}^{\prime}%
}\nonumber\\
&\qquad +  \sum_{\mathbf{k}}\left(  \epsilon_{\mathbf{k}}+\epsilon_{\mathbf{k}}%
^{I}-\frac{\mathbf{k\cdot p}}{m_{I}}\right)  \hat{a}_{\mathbf{k}}^{\dag}%
\hat{a}_{\mathbf{k}}+\hat{H}_\text{int}. \label{H 2}%
\end{align}
As in the Bose case, the LLP Hamiltonian (\ref{H 2}) distinguishes itself with
a quartic interaction term,
\begin{equation}
\hat{H}_\text{int}=\frac{1}{2m_{I}}\sum_{\mathbf{k},\mathbf{k}^{\prime}}\left(
\mathbf{k}\cdot\mathbf{k}^{\prime}\right)  \hat{a}_{\mathbf{k}}^{\dag}\hat
{a}_{\mathbf{k}^{\prime}}^{\dag}\hat{a}_{\mathbf{k}^{\prime}}\hat
{a}_{\mathbf{k}}, \label{H_int}%
\end{equation}
which indicates that a pair of host particles with momenta
$\mathbf{k}$ and $\mathbf{k}^{\prime}$, which are non-interacting in the lab frame, interact in the moving frame with an interaction potential
linearly proportional to $\mathbf{k}\cdot\mathbf{k}^{\prime}$ but inversely
proportional to the impurity mass $m_{I}$.

Usually systems described by the LLP Hamiltonian, owing to $\hat{H}_\text{int}$ being quartic, cannot be solved exactly. Thankfully there exist a
rich set of field theoretic techniques for solving (approximately) many-body quantum systems
containing four-fermion interactions \cite{hewson97HeavyFermionsBook}.  In this paper, we circumvent complications arising from the LLP induced Fermi-Fermi interaction (\ref{H_int}) by adapting the Hartree-Fock (HF) treatment to the LLP Hamiltonian in Eq.\ (\ref{H 2}) at zero temperature.  Note that the $T$-matrix approximation involving the use of ladder diagrams, a field theoretical technique, was used in the study of polarons described by the original Hamiltonian (\ref{H prime})
\cite{combescot07PhysRevLett.98.180402}, but there the use of the technique was made for a very different reason---to combat difficulties arising from the
impurity-fermion interaction which is quartic in the original Hamiltonian (\ref{H prime}).

The essence of the HF approach is summarized as follows. For a free Fermi gas, the many-body ground state (Fermi sea) is a Slater determinant of the lowest single-particle momentum states $| \mathbf{k}\rangle $
up to the Fermi energy $\epsilon_{F}=k_{F}^{2}/2m$ with $k_{F}$ the
Fermi momentum. The HF approximation amounts to assuming that for an
interacting Fermi gas, the many-body ground state $| \phi
\rangle $ continues to be in a Slater determinant but composed of a set
of orthonormal single-particle orbitals $| n\rangle $, known as
Hartree-Fock orbitals, up to the chemical potential $\mu$. 

Edwards proposed this ansatz in the position representation
\cite{Edwards1990PThPS.101.453}, inspired by the wave function introduced by Wigner and
Seitz in their discussion of electron correlations in sodium
\cite{wigner34PhysRev.46.509}. Instead, we formulate the theory in a manner
parallel to our recent work for Fr\"{o}hlich polarons
\cite{kain16PhysRevA.94.013621} except that pair correlations, which are
important in Bose polaron systems
\cite{grusdt15ScientificReports.5.12124,shchadilova16PhysRevA.93.043606,kain16PhysRevA.94.013621}, are excluded because of the lack of superfluidity in our Fermi model.  Note that our Hamiltonian in Eq.\ (1) is for a wide Feshbach resonance characterized by a single parameter---the $s$-wave scattering length.  As such our HF ansatz can describe the polaron-molecule crossover; it cannot capture three-body physics where parameters besides the $s$-wave scattering length are required to describe the impurity-host fermion interaction.  

As with our earlier study \cite{kain16PhysRevA.94.013621}, instead of  single-particle orbitals, we find it more convenient to work with the (Hermitian) single-particle
density matrix $\rho$ associated with $| \phi
\rangle $, defined as
\begin{equation}
\rho_{\mathbf{kk}^{\prime}}=\left\langle \phi\right\vert \hat{a}%
_{\mathbf{k}^{\prime}}^{\dag}\hat{a}_{\mathbf{k}}\left\vert \phi\right\rangle.
\end{equation}
The average energy, $E\equiv\langle \phi | \hat{H} |
\phi\rangle $, for a system prepared in state $|\phi\rangle$ is then a functional of $\rho$ given by
\begin{align}
E  &  =\frac{p^{2}}{2m_{I}}+\sum_{\mathbf{k}}\left(  \epsilon_{\mathbf{k}%
}+\epsilon_{\mathbf{k}}^{I}-\frac{\mathbf{k\cdot p}}{m_{I}}\right)
\rho_{\mathbf{kk}}\nonumber\\
&\qquad+  \frac{g}{\mathcal{V}}\sum_{\mathbf{k},\mathbf{k}^{\prime}}\rho
_{\mathbf{kk}^{\prime}}+\sum_{\mathbf{k},\mathbf{k}^{\prime}}\frac
{\mathbf{k}\cdot\mathbf{k}^{\prime}}{2m_{I}}\left(  \rho_{\mathbf{kk}}%
\rho_{\mathbf{k}^{\prime}\mathbf{k}^{\prime}}-\left\vert \rho_{\mathbf{kk}%
^{\prime}}\right\vert ^{2}\right)  , \label{E}%
\end{align}
where the term proportional to $\rho_{\mathbf{kk}}\rho_{\mathbf{k}^{\prime
}\mathbf{k}^{\prime}}$ is due to the Hartree contribution and the  $-\left\vert
\rho_{\mathbf{kk}^{\prime}}\right\vert ^{2}$ term follows from the Fock exchange
contribution.  Note that for $\rho$ to represent a Slater determinant it must satisfy
\begin{equation}
\rho^{2}=\rho. \label{constraint for rho}%
\end{equation}

Minimizing $E$ in Eq.\ (\ref{E}) with respect to $\rho$ subject to
condition (\ref{constraint for rho}), i.e.
\begin{equation}
\delta\left(  E-Tr\left[  \Lambda\left(  \rho^{2}-\rho\right)  \right]
\right)  =0,
\end{equation}
we arrive at the HF equation,
\begin{equation}
\left[  A,\rho\right]  =0, \label{HF equation}
\end{equation}
where $\Lambda$ is a matrix of Lagrange multipliers associated with
constraint (\ref{constraint for rho}) and $A$ is a matrix defined as
$A_{\mathbf{kk}^{\prime}}=\partial E/\partial\rho_{\mathbf{k}^{\prime
}\mathbf{k}}$ or explicitly
\begin{equation}
A_{\mathbf{kk}^{\prime}}=\left[  \epsilon_{\mathbf{k}}+\epsilon_{\mathbf{k}%
}^{I}-\frac{\mathbf{k\cdot}\left(  \mathbf{p}-\mathbf{p}_{f}\right)  }{m_{I}%
}\right]  \delta_{\mathbf{k},\mathbf{k}^{\prime}}+\frac{g}{\mathcal{V}}%
-\frac{\mathbf{k}\cdot\mathbf{k}^{\prime}}{m_{I}}\rho_{\mathbf{kk}^{\prime}},
\label{A_kk'}%
\end{equation}
with
\begin{equation}
\mathbf{p}_{f}=\sum_{\mathbf{k}}\mathbf{k}\left\langle \phi\right\vert \hat
{a}_{\mathbf{k}}^{\dag}\hat{a}_{\mathbf{k}}\left\vert \phi\right\rangle
=\sum_{\mathbf{k}}\mathbf{k}\rho_{\mathbf{kk}}
\end{equation}
being the total fermion momentum. 

The HF equation (\ref{HF equation}) is automatically satisfied when one
chooses the HF orbital $\left\vert n\right\rangle $ as the eigenstate of $A$:
\begin{equation}
A\left\vert n\right\rangle =\omega_{n}\left\vert n\right\rangle ,
\label{A|n>=}%
\end{equation}
where $\omega_{n}$ is the eigenvalue and
\begin{equation}
\left\vert n\right\rangle =\sum_{\mathbf{k}}U_{\mathbf{k}n}\left\vert
\mathbf{k}\right\rangle , \label{U}%
\end{equation}
is the corresponding eigenstate normalized according to
\begin{equation}
\sum_{\mathbf{k}}\left\vert U_{\mathbf{k}n}\right\vert ^{2}=1.
\end{equation}
The density matrix $\rho$ is then constructed as the projector onto
the space spanned by occupied orbitals $\{  | n\rangle\}  $:
\begin{equation}
\rho=\sum_{n}\left\vert n\right\rangle \left\langle n\right\vert \theta\left(
\mu-\omega_{n}\right)  , \label{rho |n><n|}%
\end{equation}
where the step function $\theta(  \mu-\omega_{n})  $ is introduced
to impose the Pauli exclusion principle at zero temperature.  Equation
(\ref{A|n>=}) is the momentum space representation of Edward's position space HF equations \cite{Edwards1990PThPS.101.453}.

In the actual numerical implementation, the single-particle HF orbital
$| n\rangle $ has to be determined iteratively
(starting, in our implementation, by assuming fermions in a Fermi sea).  This is because Eq.\
(\ref{A|n>=}) is a nonlinear equation; matrix $A$ is itself a function of
$| n\rangle $ via the density matrix element in Eq.\
(\ref{A_kk'}),
\begin{equation}
\rho_{\mathbf{kk}^{\prime}}=\sum_{n}U_{\mathbf{k}n}^{\ast}U_{\mathbf{k}%
^{\prime}n}\theta\left(  \mu-\omega_{n}\right),
\end{equation}
where the chemical potential $\mu$ is fixed by the fermion number conservation
law,
\begin{equation}
n_F=\frac{1}{\mathcal{V}}\sum_{\mathbf{k}}\rho_{\mathbf{kk}},
\end{equation}
with $n_F$  the background fermion number density.  This simple iterative procedure can only lead to the ground state and therefore cannot capture phenomena associated with excited states such as the Fermi super-Tonks state \cite{Chen Guan}.

The HF method amounts to moving from $\{ | \mathbf{k}
\rangle \}  $ space where $A$ has the matrix representation
(\ref{A_kk'}) to $\{ | n\rangle \}  $ space where
$A$ is diagonalized through a unitary transformation $U$, with $U_{\mathbf{k}n}=\left\langle \mathbf{k}| n\right\rangle$ defined in Eq.\ (\ref{U}).  This unitary transformation induces a linear mapping
between field operators in the two spaces,
\begin{equation}
\hat{d}_{n}=\sum_{\mathbf{k}}\hat{a}_{\mathbf{k}}U_{\mathbf{k}n}, \label{d_n}
\end{equation}
which defines the quasiparticle field operator $\hat{d}_{n}$ in $\{
| n \rangle \} $ space.  The many-body HF state
$| \phi \rangle $, when viewed in $\{  |
n \rangle \} $ space, contains neither quasiparticles nor
quasiholes, i.e.\ it is a vacuum with respect to both $\hat{d}_{n}$ with $\omega_{n}>\mu$
and $\hat{d}_{n}^{\dag}$ with $\omega_{n}<\mu$. But, when viewed in $\{
| \mathbf{k}\rangle \}  $ space $|
\phi\rangle $ is a superposition of terms with different numbers of
particle-hole pair excitations (relative to the Fermi sea in $\{
| \mathbf{k}\rangle \}  $ space).  In contrast to the
trial state expansion in the weak coupling theory which includes terms only up
to a limited number of particle-hole pairs, the trial state $|
\phi\rangle $ in the HF theory, when expanded, consists of 
an unrestricted number of particle-hole pairs as illustrated in Fig.\
\ref{Fig: variationalAnsatz}(c).  In fact, the HF equation adjusts
particle-hole pairs by itself according to the impurity-fermion interaction
strength and the impurity-fermion mass ratio.  Thus, the
self-consistent HF theory is a nonperturbative variational approach that is
expected to perform well even in the limit of strong impurity-fermion
coupling.
\begin{figure}
\centering
\includegraphics[width=3.25 in]{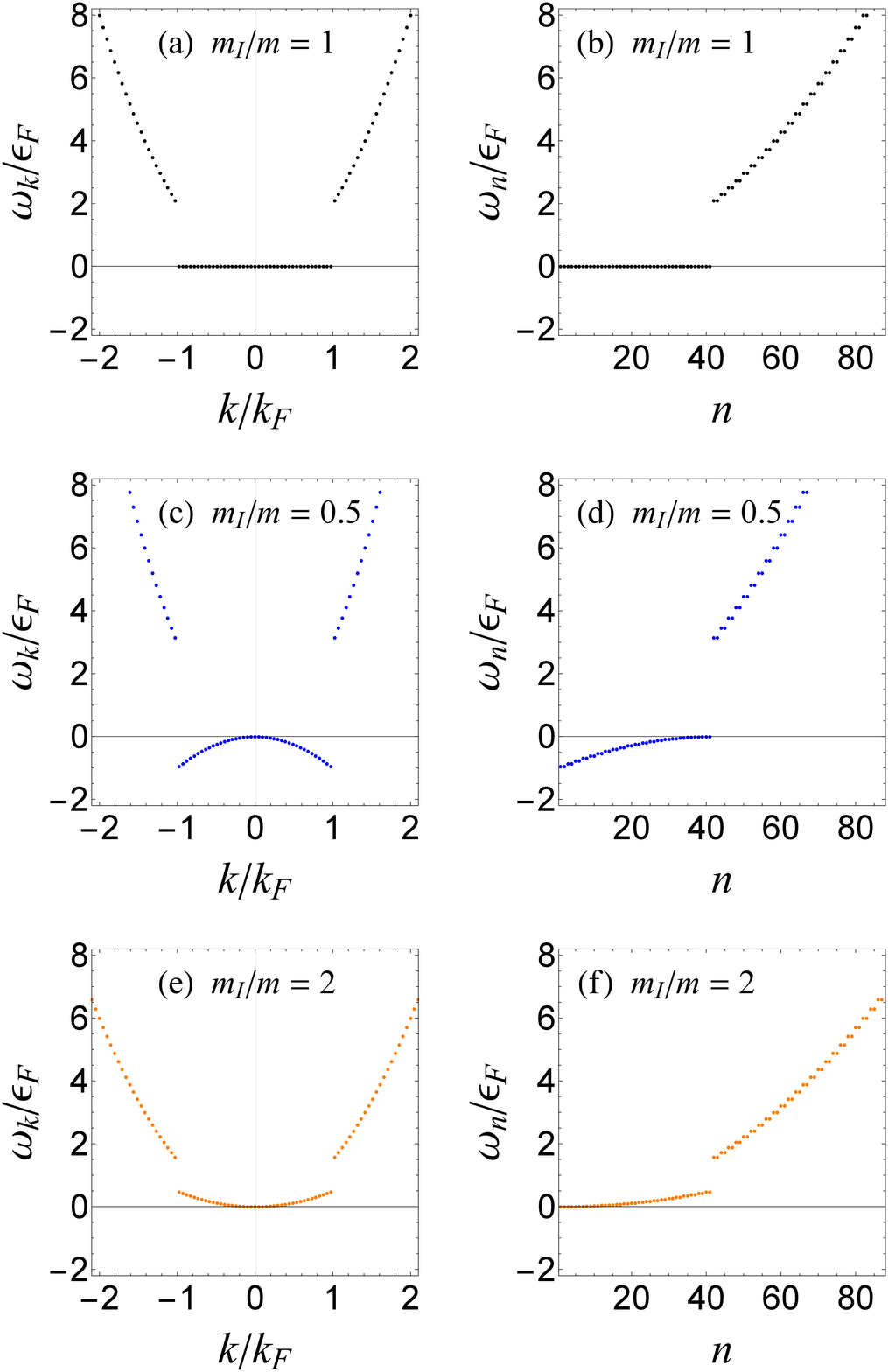}
\caption{(Color online.) The HF single-particle energy spectrum for free fermions in a one-dimensional Fermi sea ($g = p = 0$) as seen in the moving (LLP) frame as a function
of $k$ (left column) and as a function of energy quantum number $n$ (right column) when $m_I/m = 1$ (top row, black), $0.5$ (middle row, blue), and 2 (bottom row, orange).  These and all subsequent figures were made with a momentum cutoff of $25 k_F$.}
\label{Fig:w_n}
\end{figure}

The polaron energy (at momentum $\mathbf{p}$) is the total energy $E$ in Eq.\
(\ref{E}) relative to the energy of the Fermi sea, $\sum_{\mathbf{k}}%
\epsilon_{\mathbf{k}}\theta\left(  \epsilon_{F}-\epsilon_{\mathbf{k}}\right)
$, which simplifies when Eqs.\ (\ref{A|n>=}) and (\ref{rho |n><n|}) are taken
into consideration to
\begin{equation}
E_{p}=\mathcal{E}_{0}+\sum_{n}\omega_{n}\theta\left(  \mu-\omega_{n}\right)
-\sum_{\mathbf{k}}\epsilon_{\mathbf{k}}\theta\left(  \epsilon_{F}%
-\epsilon_{\mathbf{k}}\right)  , \label{E_P}%
\end{equation}
where%
\begin{equation}
\mathcal{E}_{0}=\frac{p^{2}}{2m_{I}}-\frac{p_{f}^{2}}{2m_{I}}%
+\frac{1}{2}\sum_{\mathbf{k},\mathbf{k}^{\prime}}\frac{\mathbf{k}%
\cdot\mathbf{k}^{\prime}}{m_{I}}\left\vert \rho_{\mathbf{kk}^{\prime}%
}\right\vert ^{2}. \label{H0}%
\end{equation}

For $\mathbf{p}$ near zero, Eq.\ (\ref{E_P}) takes the approximate form
$E_{p}\approx E_{0}+p^{2}/2m_{I}^{\ast}$, where $E_{0}$ is the ground state polaron
energy ($p=0$) and
\begin{equation}
m_{I}^{\ast}=\left.  \frac{\partial^{2}E_{p} }{\partial
p^{2}}\right\vert _{p=0}%
\end{equation}
is defined as the effective impurity mass, which, because of momentum
conservation, is equivalent to%
\begin{equation}
m_{I}^{\ast}=\left(  \frac{1}{m_{I}}-\frac{1}{m_{I}}\lim_{p\rightarrow0}%
\frac{\mathbf{p}_{f}\cdot\mathbf{p}}{p^{2}}\right)  ^{-1}. \label{m_I*}%
\end{equation}

The central task of the HF approach is to solve Eq.\ (\ref{A|n>=}) self
consistently for single-particle eigenenergies and corresponding eigenvectors.  The single-particle energy spectrum in the moving (LLP) frame of
reference (and within the HF approximation), as determined from Eq.\ (\ref{A|n>=}), is expected to look
drastically different from that in the lab frame. As
preparation for a detailed study in the next section, we conclude this section
using the Fermi sea as an example to clarify this point. The Fermi sea
in the lab frame is filled with fermions with free particle dispersion
$\epsilon_{\mathbf{k}}$. In contrast, the same Fermi sea in the moving
frame is occupied by fermions with a quite different energy dispersion:
\begin{equation}
\omega_{\mathbf{k}}=\left\{
\begin{array}{ll}
\epsilon_{\mathbf{k}}-\epsilon_{\mathbf{k}}^{I} & \text{ if } \left\vert
\mathbf{k}\right\vert <k_{F}\\
\epsilon_{\mathbf{k}}+\epsilon_{\mathbf{k}}^{I} & \text{ if } \left\vert
\mathbf{k}\right\vert >k_{F},
\end{array}
\right.   \label{w_n}%
\end{equation}
which we obtained by describing the Fermi sea using density matrix $\rho_{\mathbf{kk}^{\prime}}=\delta
_{\mathbf{k,k}^{\prime}}\theta(  \epsilon_{F}-\epsilon_{\mathbf{k}%
})  $.  This is shown in the left column of Fig.\ \ref{Fig:w_n}, where an energy jump of width $2
\epsilon _{k_{F}}^{I}=k_{F}^{2}/m_{I}$ at $k=\pm k_{F}$ can be seen.  Only in the heavy impurity limit, $m_{I}=\infty$, does this discontinuity vanish and do the single-particle dispersions in the lab and moving frames agree.  The energy spectrum in $\{ |n\rangle \}$ space is shown in the right column in Fig.\ \ref{Fig:w_n} where the energy jump at the Fermi surface has a width $2\epsilon_{k_{F}}^{I}$ for $m_{I}>m$ and $\epsilon _{k_{F}}+\epsilon _{k_{F}}^{I}$ for $m_{I}<m$.


\section{Discussion:\ 1D system}

In this section we specialize the HF polaron theory to a quasi-1D
(1D) system where sufficiently high harmonic trap potentials along the
transverse dimension are employed to confine the motion of atoms to zero-point
oscillations. The effective 1D coupling constant $g$ can be tuned from
negative to positive via confinement-induced resonance
\cite{olshanii98PhysRevLett.81.938,bergeman03PhysRevLett.91.163201}. The
precise value of $g$ in 1D can be obtained from the corresponding 3D
scattering length following well-established recipes, regardless of whether
the impurity and host fermions experience the same
\cite{olshanii98PhysRevLett.81.938} or different
\cite{peano05NewJournalOfPhysics.7.192} trap frequencies.  Not only is this
1D model of fundamental importance in its own right (see, for example,
\cite{Guan13RevModPhys.85.1633} for a review), but it also affords us
a proof-of-principle opportunity to test our HF theory.  First, the 1D model, unlike its 2D and 3D counterparts, does not suffer from ultraviolet divergences and therefore is not in need of being renormalized before application of our theory.  Second, there exists an exact analytical solution
due to McGuire \cite{McGuire65JMathPhys.6.432,McGuire66JMathPhys.7.123} for
the case of equal masses $m_{I}=m$.  Third, a detailed study based on Chevy's
ansatz is available in the literature \cite{giraud09PhysRevA.79.043615}, which
serves as another important benchmark against which our approach can be compared.

The Hartree-Fock ansatz can accommodate both positive and negative $g$. In order to focus on the role of particle-hole excitations in polaron physics, in what follows we limit our study to models with a (strong) positive $g$.  Models with a (strong) negative $g$ are known to be dominated by a bound state \cite{giraud09PhysRevA.79.043615} whose study we leave for future work.”


\subsection{Equal Masses:\ $m_{I}=m$}

Consider first the case where the impurity and host fermion masses are
equal, $m_{I}=m$, which, according to Eq.\ (\ref{w_n}), features a Fermi sea where all occupied states are degenerate with zero energy and the discontinuity at the Fermi surface has a width twice as large as the Fermi
energy $\epsilon_{F}$, as illustrated in the $m_{I}/m=1$ curve in Fig.\
\ref{Fig:w_n}(a).  Figure \ref{Fig:w_n}(b) shows the
single-particle spectrum in $\{| n\rangle \}$ space where it has a two-fold degeneracy due to inversion symmetry.
Note that no states exist inside the discontinuity.

\begin{figure}[ptb]
\centering
\includegraphics[
width=3.2in
]%
{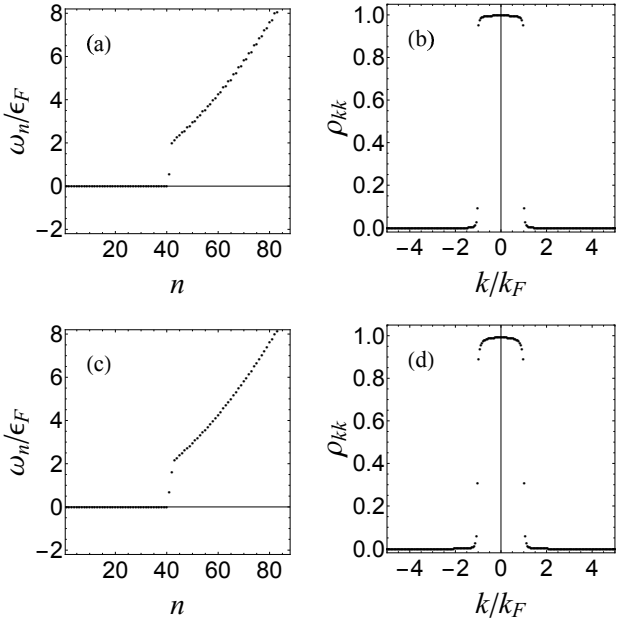}%
\caption{The single-particle energy spectrum $\omega_{n}$ (left column) and momentum distribution $\rho_{\mathbf{kk}}$ (right column) for equal masses $m_I/m = 1$ when $gm/2k_F = 1$ (top row) and $5$ (bottom row).}
\label{Fig:w_n spectrum}%
\end{figure}

If we increase $g$ we see in Fig.\ \ref{Fig:w_n spectrum}(a) that the state
directly below the discontinuity (i.e.\ directly below the Fermi surface) breaks away from the
Fermi sea and enters the gap.  The chemical potential is determined by this
break-away state, which changes from 0 (when $g=0$) to a finite value proportional to $g$.  In
addition, the two-fold degeneracy  for $g=0$ is lifted by the
anisotropy of the effective two-body interaction (\ref{H_int}).  If we
increase $g$ further we see in Fig.\ \ref{Fig:w_n spectrum}(c) that the state
directly above the discontinuity now breaks away and joins the state that broke
away from below, forming an energy spectrum characterized by two
``in-gap" states.

Important to Fermi polaron physics is the level of particle-hole activity.
 Focus now on the momentum distribution in Figs.\ \ref{Fig:w_n spectrum}(b)
and (d).  We see that increasing the impurity-fermion interaction $g$ tends
to reduce the sharpness of $\rho_{\mathbf{kk}}$ and hence, as expected,
increase the particle-hole excitations near the Fermi surface.

These features of the energy spectrum $\omega_{n}$ (i.e.\ two ``in-gap" states) and the momentum distribution $\rho
_{\mathbf{kk}}$ are found to remain essentially the same as $g$ is further
increased, implying that the system saturates, which is known to occur in 1D models
in the strong repulsive limit $g\rightarrow +\infty$
\cite{giraud09PhysRevA.79.043615}. 

McGuire, using the Bethe ansatz (BA) \cite{Bethe1931Zphys.71.205}, showed that
the Fermi polaron model with equal masses is integrable
\cite{McGuire65JMathPhys.6.432,McGuire66JMathPhys.7.123}.  McGuire's work
was the catalyst that led Yang \cite{yang67PhysRevLett.19.1312} and
Gaudin \cite{gaudin67PhysLettA.24.55} to the exact solution for 1D Fermi gases
and Lieb and Wu \cite{lieb68PhysRevLett.20.1445} to the exact solution
of the 1D Hubbard model with arbitrary spin population imbalance. McGuire's
treatment also motivated Edwards \cite{Edwards1990PThPS.101.453} to expand the HF orbital $\left\vert
n\right\rangle $ for a system of $N$ fermions in terms of $N+1$ plane-wave
states $\left\vert k_{t}\right\rangle $:
\begin{equation}
\left\vert n\right\rangle =\sum_{t=0}^{N}a_{nt}\left\vert k_{t}\right\rangle,
\quad
n=1,2,\ldots,N,
\end{equation}
where the momenta $k_{t}$ are given by the BA-like equations
\begin{equation}
\mathcal{V}k_{t}=2\pi n_{t}-2\delta_{t}%
\end{equation}
and
\begin{equation}
\delta_{t}=-\pi sgn\left(  k_{t}\right)  /2+\tan^{-1}\left[  \left(
2k_{t}-\Lambda\right)  /mg\right]  ,
\end{equation}
with $n_{t}$ an integer and $\Lambda$ the spectral parameter fixed by the
total momentum $p=\sum_{t=0}^{N}k_{t}$. Expanding the polaron energy
\begin{equation}
E_p = \sum_{t=0}^N \frac{k_t^2}{2m} - \sum_\mathbf{k} \epsilon_\mathbf{k} \theta \left(\epsilon_F - \epsilon_\mathbf{k} \right)
\end{equation}
up to second order in $p$ leads to McGuire's
analytical results for the polaron energy
\begin{equation} \label{McGuire energy}
E_{0}=\frac{2\epsilon_{F}}{\pi}\left[  \bar{g}-\frac{\pi}{2}\bar{g}%
^{2}+\left(  1+\bar{g}^{2}\right)  \tan^{-1}\bar{g}\right]
\end{equation}
and the effective mass
\begin{equation} \label{McGuire mass}
m_{I}^{\ast}=m\frac{\left(  1-\frac{2}{\pi}\tan^{-1}\bar{g}\right)  ^{2}%
}{1-\frac{2}{\pi}\left(  \tan^{-1}\bar{g}+\frac{\bar{g}}{1+\bar{g}^{2}%
}\right)  },
\end{equation}
where $\bar{g} = gm/2k_{F}$ is a unitless quantity measuring the interaction strength.

\begin{figure}[ptb]
\centering
\includegraphics[
width=3.4in
]{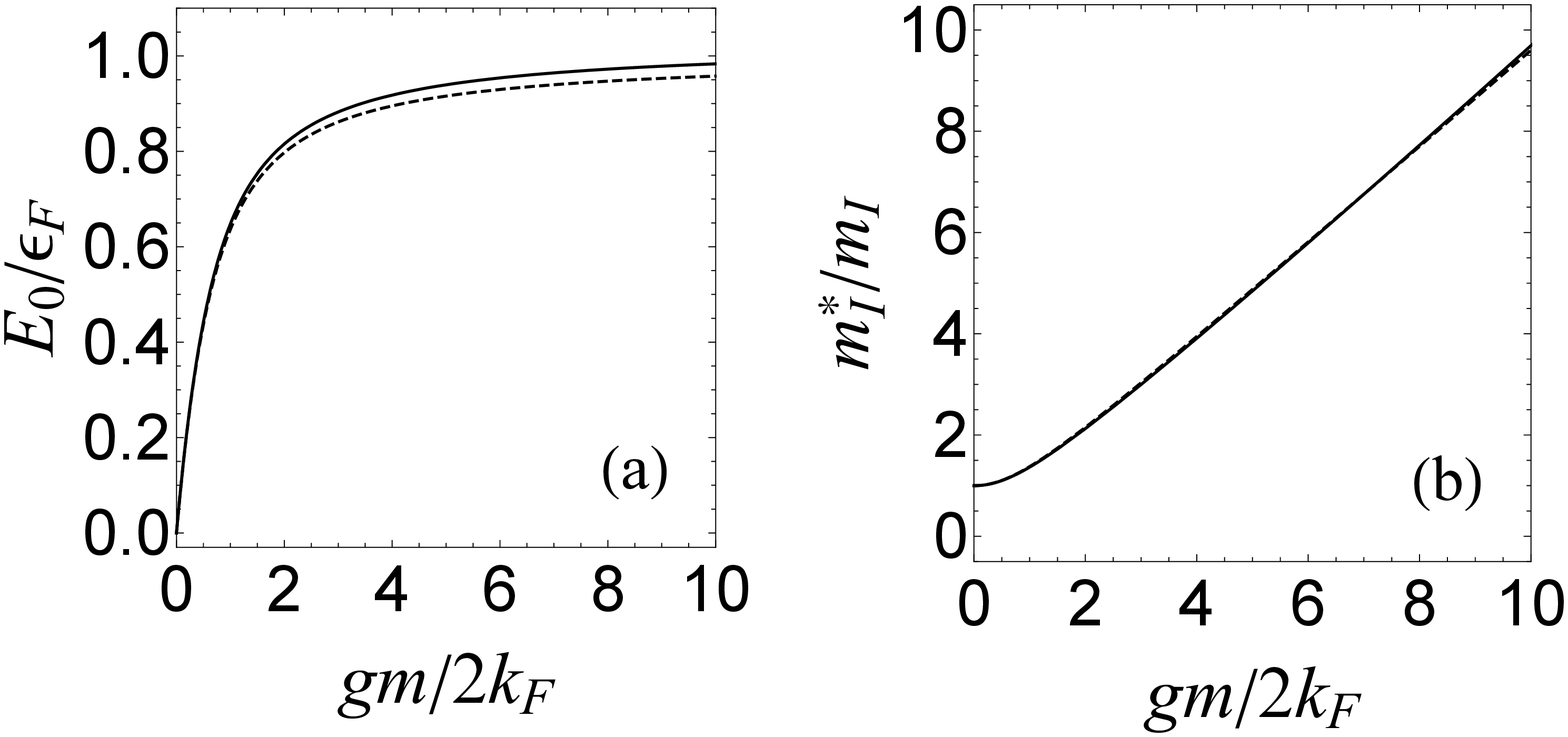}%
\caption{(a) Polaron energy $E_{0}$ and (b) effective polaron mass $m_{I}^{\ast}$, both as functions of $g$ and for equal masses $m_I = m$ (and for $p=0$).  Solid curves are from our HF theory and dotted curves are McGuire's exact results.}
\label{Fig:E_P repulsive}
\end{figure}

McGuire's exact analytical results for $m_I = m$, then, afford an excellent opportunity for to test our model. Figures \ref{Fig:E_P repulsive}(a) and \ref{Fig:E_P repulsive}(b) display, respectively, the polaron energy and effective polaron mass as functions of the impurity-fermion interaction strength $g$. The results computed using our HF theory (solid lines) are seen to be in remarkable agreement with McGuire's exact analytical solutions (dashed lines).  That our polaron energy is slightly higher than McGuire's is not unexpected since our ansatz is a variational one---so that we may apply our theory to situations with $m_I \neq m$, where states similar to $|k_t\rangle$ do not exist, we expanded the HF orbital in Eq.\ (\ref{U}) in terms of the usual free-particle states whose wave vector $k$ is different from $k_{t}$ used in the BA equations which is modified by the impurity-fermion scattering.


\subsection{Arbitrary Mass Impurity}

Having conducted a detailed comparison for the equal mass case, we now turn
our attention to the cases where the impurity mass is not equal to the host
particle mass, $m_{I}\neq m$. 

\begin{figure}[ptb]%
\centering
\includegraphics[
width=3.4in
]
{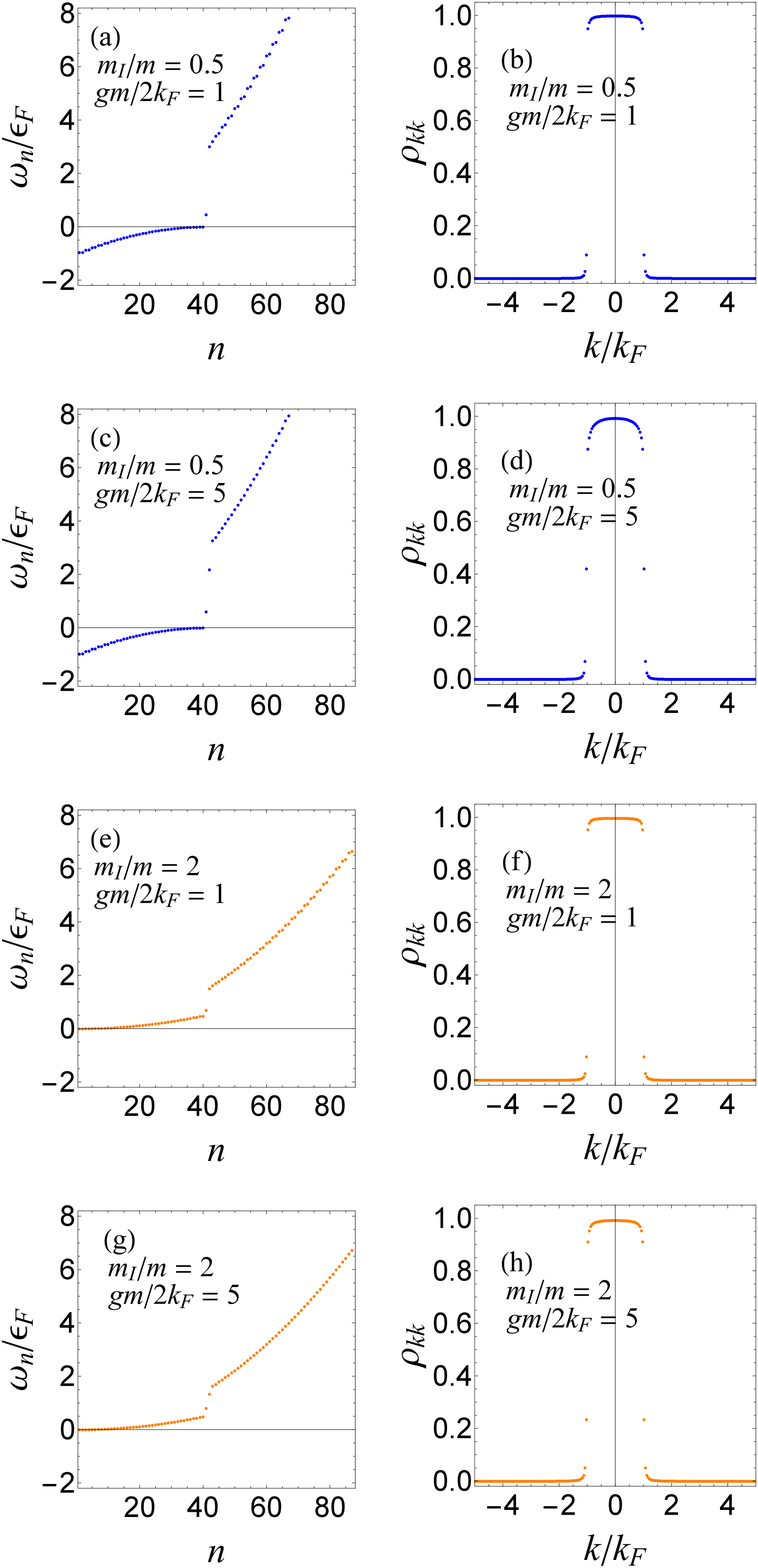}
\caption{HF single-particle energy spectrum $\omega_{n}$ (left column) and momentum
distribution $\rho_{\mathbf{kk}}$ (right column) for $m_{I}/m= 0.5$ (top two rows) and $2$ (bottom two rows) and $gm/2k_F = 1$ (first and third rows) and 5 (second and fourth rows).}
\label{Fig:heavy}
\end{figure}

Figure \ref{Fig:heavy} displays typical single-particle spectra (left column)
and momentum distributions (right column) for a lighter impurity $m_{I}<m$
(top two rows) and for a heavier impurity $m_{I}>m$ (bottom two rows).

Below the Fermi surface, the light impurity spectrum is negative and concave
while the heavy impurity spectrum is positive and convex.  At the Fermi
surface, the former exhibits a larger jump while the latter exhibits a smaller
jump compared to the equal mass case in Fig.\ \ref{Fig:w_n spectrum}.  The
features inside the discontinuity remains qualitatively the same as those for
equal masses, i.e.\ as $g$ increases there emerges first one and then two
isolated states.

The momentum distributions in the right column in Fig.\ \ref{Fig:heavy} agree with
our physical intuition based on the LLP  induced Fermi-Fermi
interaction (\ref{H_int}), namely, for a given $g$ a heavier impurity
suppresses while a lighter impurity enhances particle-hole activities compared
to the equal mass case in Fig.\ \ref{Fig:w_n spectrum}.

\begin{figure}[ptb]%
\centering
\includegraphics[
width=3.4in
]%
{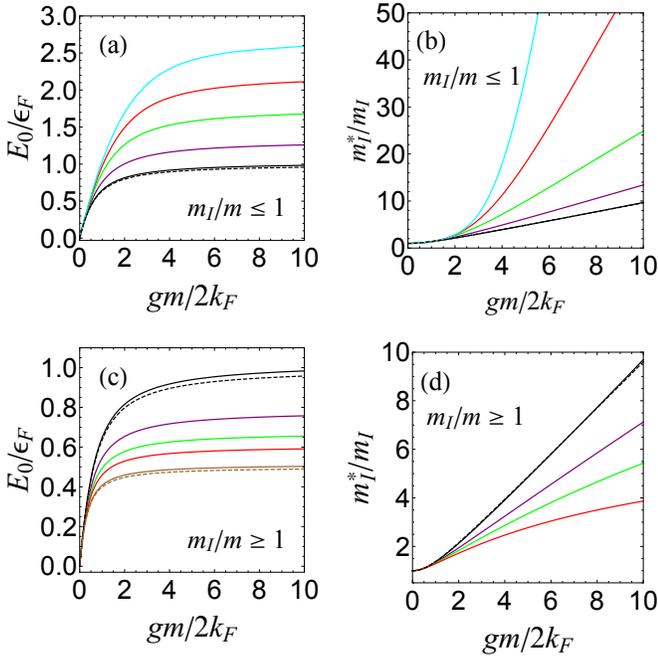}%
\caption{(Color online.) Polaron energy $E_{0}$ (first column) and effective mass $m_{I}^{\ast}$ (second column).  The top row is for an impurity mass lighter than or equal to the host fermion mass.  From bottom to top $m_I/m = 1$ (black), 0.5 (purple), 0.25 (green), 0.15 (red), and 0.1 (cyan).  The dashed lines are McGuire's exact results in Eqs.\ (\ref{McGuire energy}) and (\ref{McGuire mass}).  The bottom row is for an impurity mass heavier or equal to the host fermion mass.  From top to bottom $m_I/m = 1$ (black), 2.5 (purple), 5 (green), and 10 (red).  The bottom (brown) curve in (a) is the polaron energy for $m_I/m = \infty$.  Dashed lines are again exact results, Eqs.\ (\ref{McGuire energy}) and (\ref{McGuire mass}) for $m_I/m=1$ and Eq.\ (\ref{E infinite}) for $m_I/m = \infty$.}
\label{Fig:PolaronEnergyAndMassArbitraryCase}
\end{figure}

Figure \ref{Fig:PolaronEnergyAndMassArbitraryCase} shows the polaron energies
and effective masses as functions of $g$ for lighter impurities (top row) and
heavier impurities (bottom row). For large $g$ the polaron energy plateaus
while the effective mass continues to increase, just as in the equal mass
case. 

For small $g$, $E_0$ and $m_I^*$ are found to increase according to
perturbation theory in which $\hat{V}$ in Eq.\ (\ref{H_S}) is treated as a
perturbation to $\hat{H}_{0}$ in Eq.\ (\ref{H_0}).  Let  $E_0^{(
0)}$ and $E_{\mathbf{kq}}^{(1)}$ be, respectively, the
eigenenergies of the eigenstates $| 0\rangle $ and $|1_{\mathbf{k}}1_{\mathbf{q}}\rangle $ of $\hat{H}_{0}$.  The polaron
energy $E_{p}$ up to second order in $g$ is given by
\begin{equation}
E_{p}= \frac{p^{2}}{2m_{I}}+\left\langle 0\right\vert \hat{V}\left\vert 0\right\rangle
+
\sum_{\substack{\mathbf{q}<k_{F} \\ \mathbf{k}>k_{F}}}
\frac{\left\vert \left\langle
0\right\vert \hat{V}\left\vert 1_{\mathbf{k}}1_{\mathbf{q}}\right\rangle
\right\vert ^{2}}{E_{0}^{\left(  0\right)  }-E_{\mathbf{kq}}^{\left(
1\right)  }}%
\end{equation}
or explicitly\ \
\begin{align}
E_{p}  &  =\frac{p^{2}}{2m_{I}}+gn_{F}-2\frac{g^{2}m_{I}}{\left(  2\pi\right)
^{2}}\nonumber\\
&\qquad \times  \int\int\frac{dqdk}{q-k}\frac{1}{2p+q-k-\frac{m_{I}}{m}\left(  q+k\right)
},
\end{align}
which yields the polaron energy
\begin{equation}
E_{0}=\frac{4}{\pi}\epsilon_{F}\bar{g}\left[  1-\frac{\bar{g}}{\pi}\left(
\frac{\pi^{2}}{4}+\text{Li}_{2}\left(  \alpha\right)  -\text{Li}_{2}\left(  -\alpha\right)
\right)  \right]  \label{E_0 perturbation}%
\end{equation}
\cite{giraud09PhysRevA.79.043615} and the effective mass
\begin{equation}
\frac{m_{I}^{\ast}}{m_{I}}=1+\frac{4}{\pi^{2}}\left(  1+\alpha\right)
^{2}\bar{g}^{2} \label{effective mass perturbation}%
\end{equation}
up to second order in $g$, where $\text{Li}_{2}\left(  \alpha\right)  $ is the
dilogarithm function and $\alpha=(  m_{I}-m)  /(
m_{I}+m)  $.

The polaron energies for lighter impurities in Fig.\
\ref{Fig:PolaronEnergyAndMassArbitraryCase}(a) lie above the bottom curve for
equal masses. The lighter the impurity, the stronger the LLP
induced interaction (\ref{H_int}) and hence the higher  the polaron energy.
 In addition, the kinetic energy from impurity recoil, and hence
$\hat{H}_{0}$ in Eq.\ (\ref{H_0}), increases with $\hat{V}$ in Eq.\
(\ref{H_S}) when the impurity mass decreases.  Thus the lighter the impurity, the larger the domain $g$ upon
which the perturbative expansions in Eqs.\ (\ref{E_0 perturbation}) and
(\ref{effective mass perturbation}) become valid; these perturbative
expansions become exact in the zero impurity {mass limit.

The polaron energies for heavier impurities in Fig.\
\ref{Fig:PolaronEnergyAndMassArbitraryCase}(c) lie between the top curve for
equal masses and the bottom one for a localized impurity (i.e.\ $m_{I}
=\infty$). The polaron energy for a localized impurity arises solely
from the scattering of host fermions by a Dirac delta potential $g\delta
(  \mathbf{r})  $.  It can be found analytically
in 1D analogously to how it is done in 3D
by enclosing the impurity (at the origin) between $-R$ and $+R$ and analyzing
the phase shift in the thermodynamic limit $R\rightarrow\infty$
\cite{giraud09PhysRevA.79.043615}. This is identical to the set-up in which
one derives Fumi's theorem \cite{mahan00ManyParticlePhysicsBook},
\begin{equation}
E_{0}=-\int_{0}^{\epsilon_{F}}\frac{\eta\left(  \epsilon\right)  }{\pi
}d\epsilon.
\end{equation}
When applied to a 1D system where the phase depends on the energy $\epsilon$ according to
\begin{equation}
\eta\left(  \epsilon\right)  =-\tan^{-1}\left(  g\sqrt{\frac{m}{2\epsilon}%
}\right) ,
\end{equation}
Fumi's theorem yields the  polaron energy
\begin{equation} 
E_{0}=\frac{\epsilon_{F}}{\pi}\left[  2\bar{g}-\frac{\pi}{2}\left(  2\bar
{g}\right)  ^{2}+\left(  1+\left(  2\bar{g}\right)  ^{2}\right)  \tan
^{-1}\left(  2\bar{g}\right)  \right]  . \label{E infinite}
\end{equation}

The bottom most dashed curve in Fig.\ \ref{Fig:PolaronEnergyAndMassArbitraryCase}(c) displays Eq.\ (\ref{E infinite}). Note
that our ansatz (\ref{U}) for expanding the HF orbital $|
n\rangle $ in terms of the usual free-particle states becomes exact
(without having to resort to the HF approximation) for an infinitely heavy
impurity. The slight difference between our result and Eq.\ (\ref{E infinite})
is due solely to finite size effects.

We end this section with Table I which displays polaron energies for various impurity masses and interaction strengths computed using our self-consistent HF method.
\begin{table*}
\begin{center}
\begin{tabular}{|c|c|c|c|c|c|} 
\hline
\diagbox[height = 30pt, width = 60pt] {$m_I/m$}{$gm/2k_F$} & 0.1 & 1 & 10 & 50 & 100
\\
\hline
0.1 & 0.125  &  1.04  & 2.56  & 2.68 & 2.70 
\\
0.5 & 0.121  & 0.756  & 1.26  & 1.31 & 1.32 
\\
1 & 0.118 (0.118)  & 0.648 (0.637)  & 0.983 (0.958)  & 1.02 (0.992) & 1.02 (0.996) \\
2 & 0.116  & 0.566  & 0.801  & 0.826 & 0.829\\
10 & 0.112   & 0.459  & 0.591  & 0.604 & 0.605\\
$\infty$ & 0.110 (0.109)  & 0.408 (0.399)  & 0.503 (0.489)  & 0.512 (0.498) & 0.513 (0.499) \\
\hline
\end{tabular}
\end{center}
\caption{Various polaron energies, $E_0/\epsilon_F$, for impurities with mass $m_I$ and strength of interaction with host fermions $g$.  Numbers in parentheses are exact analytical results from Eqs.\ (\ref{McGuire energy}) and (\ref{E infinite}). \label{table}}
\end{table*}


\section{Conclusion}

The Fr\"{o}hlich Hamiltonian and the LLP transformation have played important roles in advancing our understanding of Bose polarons. We applied the same techniques to transform the Fermi polaron problem to the fermionic LLP model which contains only host fermions, but that interact with each
other.  We adapted the variational principle based on the HF approximation to the fermionic LLP Hamiltonian. We applied this HF theory, which has the
advantage of not imposing restrictions on the number of particle-hole pairs, to repulsive Fermi polarons in one dimension.  The polaron properties depend
crucially on the single-particle energy spectrum which features a jump at the Fermi surface in the moving frame.  We discovered that as the
impurity-fermion interaction increases, the spectrum changes from having one isolated state to having two isolated states inside the discontinuity.  We found that, for
the case where impurity and host fermion masses are equal, our results calculated from the HF variational ansatz are in excellent agreement with
McGuire's exact solution from the Bethe ansatz. Finally, we used the HF theory to compute the polaron energy for impurities with different masses and
interaction strengths.  For the remainder of this conclusion we highlight additional merits of our approach
to the Fermi polaron problem.

In addition to the polaron energy and effective mass, polaron systems can be characterized by various correlation functions, which can be  probed in cold atom experiments using, for example, time-of-flight \cite{ehud04PhysRevA.70.013603,simon05Nature.434.481,rom06Nature.434.481} and
radio-frequency spectrum techniques \cite{chin04Science305.20082004,kinnunen04Science305.20082004}. Describing a Fermi polaron system using a Slater determinant involving HF orbitals can significantly simplify the task of calculating these correlation functions, something known to be difficult to do using the Bethe ansatz.

A great impetus for the recent flood of activity in nonequilbirum polaron dynamics (see, for example, \cite{schmidt17arXiv:1702.08587} and references therein) is that coherent dynamics, which often changes too rapidly to be directly observed in solid state systems, can be studied in real time using techniques such as Ramsey interferometry  in cold atom systems \cite{cetina16Science.354.6308}.  The HF polaron theory we developed in this work has its nonequilibrium
analog---the time-dependent HF variational principle. Thus, we consider it a strength of our approach that it can  be adapted straightforwardly to nonequilibrium polaron problems.

Finally, casting the Fermi polaron problem in a moving frame has the advantage that many-body field theoretic tools other than HF methods may also be applied to solve the problem \cite{hewson97HeavyFermionsBook}.  In particular, it remains an interesting question whether renormalization group techniques, similar to those recently developed by Fabian et.\ al.\ \cite{grusdt15ScientificReports.5.12124, grusdt15arXiv:1510.04934, Grusdt17arXiv:1704.02605},
can be developed to improve our understanding of strongly-coupled Fermi polarons in higher dimensions.



\begin{thebibliography}{84}
\expandafter\ifx\csname natexlab\endcsname\relax\def\natexlab#1{#1}\fi
\expandafter\ifx\csname bibnamefont\endcsname\relax
  \def\bibnamefont#1{#1}\fi
\expandafter\ifx\csname bibfnamefont\endcsname\relax
  \def\bibfnamefont#1{#1}\fi
\expandafter\ifx\csname citenamefont\endcsname\relax
  \def\citenamefont#1{#1}\fi
\expandafter\ifx\csname url\endcsname\relax
  \def\url#1{\texttt{#1}}\fi
\expandafter\ifx\csname urlprefix\endcsname\relax\def\urlprefix{URL }\fi
\providecommand{\bibinfo}[2]{#2}
\providecommand{\eprint}[2][]{\url{#2}}

\bibitem{landau46ZhEkspTeorFiz.16.341}
	S.\ I.\ Pekar, Zh.\ Eksp.\ Teor.\ Fiz.\ \textbf{16} 341 (1946).

\bibitem[{\citenamefont{Landau and Pekar}(1948)}]{landau48ZhEkspTeorFiz.18.419}
\bibinfo{author}{\bibfnamefont{L.~D.} \bibnamefont{Landau}} \bibnamefont{and}
  \bibinfo{author}{\bibfnamefont{S.~I.} \bibnamefont{Pekar}},
  \bibinfo{journal}{Zh. Eksp. Teor. Fiz} \textbf{\bibinfo{volume}{18}},
  \bibinfo{pages}{419} (\bibinfo{year}{1948}).

\bibitem[{\citenamefont{Anderson}(1967)}]{anderson67PhysRevLett.18.1049}
\bibinfo{author}{\bibfnamefont{P.~W.} \bibnamefont{Anderson}},
  \bibinfo{journal}{Phys. Rev. Lett.} \textbf{\bibinfo{volume}{18}},
  \bibinfo{pages}{1049} (\bibinfo{year}{1967}),

\bibitem[{\citenamefont{Mahan}(2000)}]{mahan00ManyParticlePhysicsBook}
\bibinfo{author}{\bibfnamefont{G.~D.} \bibnamefont{Mahan}},
  \bibinfo{journal}{Kluwer Academic/Plenum Publishers, New York}
  (\bibinfo{year}{2000}).

\bibitem[{\citenamefont{Kondo}(1964)}]{kondo64ProgressOfTheoreticalPhysics.32.37}
\bibinfo{author}{\bibfnamefont{J.}~\bibnamefont{Kondo}},
  \bibinfo{journal}{Progress of Theoretical Physics}
  \textbf{\bibinfo{volume}{32}}, \bibinfo{pages}{37} (\bibinfo{year}{1964}),

\bibitem[{\citenamefont{Hewson}(1997)}]{hewson97HeavyFermionsBook}
\bibinfo{author}{\bibfnamefont{A.~C.} \bibnamefont{Hewson}},
  \bibinfo{journal}{Cambridge University Press, Cambridge, UK}
  (\bibinfo{year}{1997}).

\bibitem[{\citenamefont{Kondo and
  Soda}(1983)}]{kondo83JournalOfLowTemperaturePhysic.50.21}
\bibinfo{author}{\bibfnamefont{J.}~\bibnamefont{Kondo}} \bibnamefont{and}
  \bibinfo{author}{\bibfnamefont{T.}~\bibnamefont{Soda}},
  \bibinfo{journal}{Journal of Low Temperature Physics}
  \textbf{\bibinfo{volume}{50}}, \bibinfo{pages}{21} (\bibinfo{year}{1983}),

\bibitem[{\citenamefont{Chevy and Mora}(2010)}]{chevy10RepProgPhys.73.112401}
\bibinfo{author}{\bibfnamefont{F.}~\bibnamefont{Chevy}} \bibnamefont{and}
  \bibinfo{author}{\bibfnamefont{C.}~\bibnamefont{Mora}},
  \bibinfo{journal}{Rep. Prog. Phys.} \textbf{\bibinfo{volume}{73}},
  \bibinfo{pages}{112401} (\bibinfo{year}{2010}).

\bibitem[{\citenamefont{Massignan et~al.}(2014)\citenamefont{Massignan,
  Zaccanti, and Bruun}}]{massignan14ReportsOnProgressInPhysics.77.034401}
\bibinfo{author}{\bibfnamefont{P.}~\bibnamefont{Massignan}},
  \bibinfo{author}{\bibfnamefont{M.}~\bibnamefont{Zaccanti}}, \bibnamefont{and}
  \bibinfo{author}{\bibfnamefont{G.~M.} \bibnamefont{Bruun}},
  \bibinfo{journal}{Reports on Progress in Physics}
  \textbf{\bibinfo{volume}{77}}, \bibinfo{pages}{034401}
  (\bibinfo{year}{2014}),

\bibitem[{\citenamefont{Schirotzek et~al.}(2009)\citenamefont{Schirotzek, Wu,
  Sommer, and Zwierlein}}]{schirotzek09PhysRevLett.102.230402}
\bibinfo{author}{\bibfnamefont{A.}~\bibnamefont{Schirotzek}},
  \bibinfo{author}{\bibfnamefont{C.-H.} \bibnamefont{Wu}},
  \bibinfo{author}{\bibfnamefont{A.}~\bibnamefont{Sommer}}, \bibnamefont{and}
  \bibinfo{author}{\bibfnamefont{M.~W.} \bibnamefont{Zwierlein}},
  \bibinfo{journal}{Phys. Rev. Lett.} \textbf{\bibinfo{volume}{102}},
  \bibinfo{pages}{230402} (\bibinfo{year}{2009}).

\bibitem[{\citenamefont{Nascimb\`ene et~al.}(2009)\citenamefont{Nascimb\`ene,
  Navon, Jiang, Tarruell, Teichmann, McKeever, Chevy, and
  Salomon}}]{nascimbene09PhysRevLett.103.170402}
\bibinfo{author}{\bibfnamefont{S.}~\bibnamefont{Nascimb\`ene}},
  \bibinfo{author}{\bibfnamefont{N.}~\bibnamefont{Navon}},
  \bibinfo{author}{\bibfnamefont{K.~J.} \bibnamefont{Jiang}},
  \bibinfo{author}{\bibfnamefont{L.}~\bibnamefont{Tarruell}},
  \bibinfo{author}{\bibfnamefont{M.}~\bibnamefont{Teichmann}},
  \bibinfo{author}{\bibfnamefont{J.}~\bibnamefont{McKeever}},
  \bibinfo{author}{\bibfnamefont{F.}~\bibnamefont{Chevy}}, \bibnamefont{and}
  \bibinfo{author}{\bibfnamefont{C.}~\bibnamefont{Salomon}},
  \bibinfo{journal}{Phys. Rev. Lett.} \textbf{\bibinfo{volume}{103}},
  \bibinfo{pages}{170402} (\bibinfo{year}{2009}).

\bibitem[{\citenamefont{Chevy}(2006)}]{chevy06PhysRevA.74.063628}
\bibinfo{author}{\bibfnamefont{F.}~\bibnamefont{Chevy}},
  \bibinfo{journal}{Phys. Rev. A} \textbf{\bibinfo{volume}{74}},
  \bibinfo{pages}{063628} (\bibinfo{year}{2006}).

\bibitem[{\citenamefont{Lobo et~al.}(2006)\citenamefont{Lobo, Recati, Giorgini,
  and Stringari}}]{lobo06PhysRevLett.97.200403}
\bibinfo{author}{\bibfnamefont{C.}~\bibnamefont{Lobo}},
  \bibinfo{author}{\bibfnamefont{A.}~\bibnamefont{Recati}},
  \bibinfo{author}{\bibfnamefont{S.}~\bibnamefont{Giorgini}}, \bibnamefont{and}
  \bibinfo{author}{\bibfnamefont{S.}~\bibnamefont{Stringari}},
  \bibinfo{journal}{Phys. Rev. Lett.} \textbf{\bibinfo{volume}{97}},
  \bibinfo{pages}{200403} (\bibinfo{year}{2006}),

\bibitem[{\citenamefont{Combescot et~al.}(2007)\citenamefont{Combescot, Recati,
  Lobo, and Chevy}}]{combescot07PhysRevLett.98.180402}
\bibinfo{author}{\bibfnamefont{R.}~\bibnamefont{Combescot}},
  \bibinfo{author}{\bibfnamefont{A.}~\bibnamefont{Recati}},
  \bibinfo{author}{\bibfnamefont{C.}~\bibnamefont{Lobo}}, \bibnamefont{and}
  \bibinfo{author}{\bibfnamefont{F.}~\bibnamefont{Chevy}},
  \bibinfo{journal}{Phys. Rev. Lett.} \textbf{\bibinfo{volume}{98}},
  \bibinfo{pages}{180402} (\bibinfo{year}{2007}).

\bibitem[{\citenamefont{Combescot and
  Giraud}(2008)}]{combescot08PhysRevLett.101.050404}
\bibinfo{author}{\bibfnamefont{R.}~\bibnamefont{Combescot}} \bibnamefont{and}
  \bibinfo{author}{\bibfnamefont{S.}~\bibnamefont{Giraud}},
  \bibinfo{journal}{Phys. Rev. Lett.} \textbf{\bibinfo{volume}{101}},
  \bibinfo{pages}{050404} (\bibinfo{year}{2008}),

\bibitem[{\citenamefont{Prokof'ev and
  Svistunov}(2008)}]{prokofev08PhysRevB.77.020408}
\bibinfo{author}{\bibfnamefont{N.}~\bibnamefont{Prokof'ev}} \bibnamefont{and}
  \bibinfo{author}{\bibfnamefont{B.}~\bibnamefont{Svistunov}},
  \bibinfo{journal}{Phys. Rev. B} \textbf{\bibinfo{volume}{77}},
  \bibinfo{pages}{020408} (\bibinfo{year}{2008}).

\bibitem[{\citenamefont{Zwierlein et~al.}(2006)\citenamefont{Zwierlein,
  Schirotzek, Schunck, and Ketterle}}]{zwierlein06Science.311.492}
\bibinfo{author}{\bibfnamefont{M.~W.} \bibnamefont{Zwierlein}},
  \bibinfo{author}{\bibfnamefont{A.}~\bibnamefont{Schirotzek}},
  \bibinfo{author}{\bibfnamefont{C.~H.} \bibnamefont{Schunck}},
  \bibnamefont{and} \bibinfo{author}{\bibfnamefont{W.}~\bibnamefont{Ketterle}},
  \bibinfo{journal}{Science} \textbf{\bibinfo{volume}{311}},
  \bibinfo{pages}{492} (\bibinfo{year}{2006}). 

\bibitem[{\citenamefont{Partridge et~al.}(2006)\citenamefont{Partridge, Li,
  Kamar, Liao, and Hulet}}]{partridge06Science.31.503}
\bibinfo{author}{\bibfnamefont{G.~B.} \bibnamefont{Partridge}},
  \bibinfo{author}{\bibfnamefont{W.}~\bibnamefont{Li}},
  \bibinfo{author}{\bibfnamefont{R.~I.} \bibnamefont{Kamar}},
  \bibinfo{author}{\bibfnamefont{Y.-a.} \bibnamefont{Liao}}, \bibnamefont{and}
  \bibinfo{author}{\bibfnamefont{R.~G.} \bibnamefont{Hulet}},
  \bibinfo{journal}{Science} \textbf{\bibinfo{volume}{311}},
  \bibinfo{pages}{503} (\bibinfo{year}{2006}). 

\bibitem[{\citenamefont{Astrakharchik and
  Pitaevskii}(2004)}]{astrakharchik04PhysRevA.70.013608}
\bibinfo{author}{\bibfnamefont{G.~E.} \bibnamefont{Astrakharchik}}
  \bibnamefont{and} \bibinfo{author}{\bibfnamefont{L.~P.}
  \bibnamefont{Pitaevskii}}, \bibinfo{journal}{Phys. Rev. A}
  \textbf{\bibinfo{volume}{70}}, \bibinfo{pages}{013608}
  (\bibinfo{year}{2004}).

\bibitem[{\citenamefont{Cucchietti and
  Timmermans}(2006)}]{cucchietti06PhysRevLett.96.210401}
\bibinfo{author}{\bibfnamefont{F.~M.} \bibnamefont{Cucchietti}}
  \bibnamefont{and}
  \bibinfo{author}{\bibfnamefont{E.}~\bibnamefont{Timmermans}},
  \bibinfo{journal}{Phys. Rev. Lett.} \textbf{\bibinfo{volume}{96}},
  \bibinfo{pages}{210401} (\bibinfo{year}{2006}).

\bibitem[{\citenamefont{Kalas and Blume}(2006)}]{kalas06PhysRevA.73.043608}
\bibinfo{author}{\bibfnamefont{R.~M.} \bibnamefont{Kalas}} \bibnamefont{and}
  \bibinfo{author}{\bibfnamefont{D.}~\bibnamefont{Blume}},
  \bibinfo{journal}{Phys. Rev. A} \textbf{\bibinfo{volume}{73}},
  \bibinfo{pages}{043608} (\bibinfo{year}{2006}).

\bibitem[{\citenamefont{Bruderer et~al.}(2008)\citenamefont{Bruderer, Bao, and
  Jaksch}}]{bruderer08EurPhysLett.82.30004}
\bibinfo{author}{\bibfnamefont{M.}~\bibnamefont{Bruderer}},
  \bibinfo{author}{\bibfnamefont{W.}~\bibnamefont{Bao}}, \bibnamefont{and}
  \bibinfo{author}{\bibfnamefont{D.}~\bibnamefont{Jaksch}},
  \bibinfo{journal}{Eur. Phys. Lett.} \textbf{\bibinfo{volume}{82}},
  \bibinfo{pages}{30004} (\bibinfo{year}{2008}).

\bibitem[{\citenamefont{Pilati and
  Giorgini}(2008)}]{pilati08PhysRevLett.100.030401}
\bibinfo{author}{\bibfnamefont{S.}~\bibnamefont{Pilati}} \bibnamefont{and}
  \bibinfo{author}{\bibfnamefont{S.}~\bibnamefont{Giorgini}},
  \bibinfo{journal}{Phys. Rev. Lett.} \textbf{\bibinfo{volume}{100}},
  \bibinfo{pages}{030401} (\bibinfo{year}{2008}),

\bibitem[{\citenamefont{Huang and
  Wan}(2009)}]{Huang09ChinesePhysicsLetters.26.080302}
\bibinfo{author}{\bibfnamefont{B.-B.} \bibnamefont{Huang}} \bibnamefont{and}
  \bibinfo{author}{\bibfnamefont{S.-L.} \bibnamefont{Wan}},
  \bibinfo{journal}{Chinese Physics Letters} \textbf{\bibinfo{volume}{26}},
  \bibinfo{pages}{080302} (\bibinfo{year}{2009}).

\bibitem[{\citenamefont{Tempere et~al.}(2009)\citenamefont{Tempere, Casteels,
  Oberthaler, Knoop, Timmermans, and Devreese}}]{tempere09PhysRevB.80.184504}
\bibinfo{author}{\bibfnamefont{J.}~\bibnamefont{Tempere}},
  \bibinfo{author}{\bibfnamefont{W.}~\bibnamefont{Casteels}},
  \bibinfo{author}{\bibfnamefont{M.~K.} \bibnamefont{Oberthaler}},
  \bibinfo{author}{\bibfnamefont{S.}~\bibnamefont{Knoop}},
  \bibinfo{author}{\bibfnamefont{E.}~\bibnamefont{Timmermans}},
  \bibnamefont{and} \bibinfo{author}{\bibfnamefont{J.~T.}
  \bibnamefont{Devreese}}, \bibinfo{journal}{Phys. Rev. B}
  \textbf{\bibinfo{volume}{80}}, \bibinfo{pages}{184504}
  (\bibinfo{year}{2009}).

\bibitem[{\citenamefont{Punk et~al.}(2009)\citenamefont{Punk, Dumitrescu, and
  Zwerger}}]{zwerger09PhysRevA.80.053605}
\bibinfo{author}{\bibfnamefont{M.}~\bibnamefont{Punk}},
  \bibinfo{author}{\bibfnamefont{P.~T.} \bibnamefont{Dumitrescu}},
  \bibnamefont{and} \bibinfo{author}{\bibfnamefont{W.}~\bibnamefont{Zwerger}},
  \bibinfo{journal}{Phys. Rev. A} \textbf{\bibinfo{volume}{80}},
  \bibinfo{pages}{053605} (\bibinfo{year}{2009}),

\bibitem[{\citenamefont{Cui and Zhai}(2010)}]{cui10PhysRevA.81.041602}
\bibinfo{author}{\bibfnamefont{X.}~\bibnamefont{Cui}} \bibnamefont{and}
  \bibinfo{author}{\bibfnamefont{H.}~\bibnamefont{Zhai}},
  \bibinfo{journal}{Phys. Rev. A} \textbf{\bibinfo{volume}{81}},
  \bibinfo{pages}{041602} (\bibinfo{year}{2010}).

\bibitem[{\citenamefont{Pilati et~al.}(2010)\citenamefont{Pilati, Bertaina,
  Giorgini, and Troyer}}]{pilati10PhysRevLett.105.030405}
\bibinfo{author}{\bibfnamefont{S.}~\bibnamefont{Pilati}},
  \bibinfo{author}{\bibfnamefont{G.}~\bibnamefont{Bertaina}},
  \bibinfo{author}{\bibfnamefont{S.}~\bibnamefont{Giorgini}}, \bibnamefont{and}
  \bibinfo{author}{\bibfnamefont{M.}~\bibnamefont{Troyer}},
  \bibinfo{journal}{Phys. Rev. Lett.} \textbf{\bibinfo{volume}{105}},
  \bibinfo{pages}{030405} (\bibinfo{year}{2010}),

\bibitem[{\citenamefont{Massignan and
  Bruun}(2011)}]{massignan11EurPhysJD.65.83}
\bibinfo{author}{\bibfnamefont{P.}~\bibnamefont{Massignan}} \bibnamefont{and}
  \bibinfo{author}{\bibfnamefont{G.~M.} \bibnamefont{Bruun}},
  \bibinfo{journal}{Eur. Phys. J. D.} \textbf{\bibinfo{volume}{65}},
  \bibinfo{pages}{83} (\bibinfo{year}{2011}).

\bibitem[{\citenamefont{Schmidt and Enss}(2011)}]{schmidt11PhysRevA.83.063620}
\bibinfo{author}{\bibfnamefont{R.}~\bibnamefont{Schmidt}} \bibnamefont{and}
  \bibinfo{author}{\bibfnamefont{T.}~\bibnamefont{Enss}},
  \bibinfo{journal}{Phys. Rev. A} \textbf{\bibinfo{volume}{83}},
  \bibinfo{pages}{063620} (\bibinfo{year}{2011}).

\bibitem{Mathy}
	J.\ M.\ Mathy, M.\ M.\ Parish, and D.\ A.\ Huse, Phys.\ Rev.\ Lett.\ \textbf{106}, 166404 (2011).

\bibitem[{\citenamefont{Guan}(2012)}]{guan12FrontierOfPhysics.7.8}
\bibinfo{author}{\bibfnamefont{X.-W.} \bibnamefont{Guan}},
  \bibinfo{journal}{Frontier of Physics} \textbf{\bibinfo{volume}{7}},
  \bibinfo{pages}{8} (\bibinfo{year}{2012}).

\bibitem[{\citenamefont{Yi and Cui}(2015)}]{yi15PhysRevA.92.013620}
\bibinfo{author}{\bibfnamefont{W.}~\bibnamefont{Yi}} \bibnamefont{and}
  \bibinfo{author}{\bibfnamefont{X.}~\bibnamefont{Cui}},
  \bibinfo{journal}{Phys. Rev. A} \textbf{\bibinfo{volume}{92}},
  \bibinfo{pages}{013620} (\bibinfo{year}{2015}),

\bibitem[{\citenamefont{Mao et~al.}(2016)\citenamefont{Mao, Guan, and
  Wu}}]{Mao16PhysRevA.94.043645}
\bibinfo{author}{\bibfnamefont{R.}~\bibnamefont{Mao}},
  \bibinfo{author}{\bibfnamefont{X.~W.} \bibnamefont{Guan}}, \bibnamefont{and}
  \bibinfo{author}{\bibfnamefont{B.}~\bibnamefont{Wu}}, \bibinfo{journal}{Phys.
  Rev. A} \textbf{\bibinfo{volume}{94}}, \bibinfo{pages}{043645}
  (\bibinfo{year}{2016}),

\bibitem[{\citenamefont{Schmidt et~al.}(2017)\citenamefont{Schmidt, Knap,
  Ivanov, You, Cetina, and Demler}}]{schmidt17arXiv:1702.08587}
\bibinfo{author}{\bibfnamefont{R.}~\bibnamefont{Schmidt}},
  \bibinfo{author}{\bibfnamefont{M.}~\bibnamefont{Knap}},
  \bibinfo{author}{\bibfnamefont{D.~A.} \bibnamefont{Ivanov}},
  \bibinfo{author}{\bibfnamefont{J.-S.} \bibnamefont{You}},
  \bibinfo{author}{\bibfnamefont{M.}~\bibnamefont{Cetina}}, \bibnamefont{and}
  \bibinfo{author}{\bibfnamefont{E.}~\bibnamefont{Demler}},
  \bibinfo{journal}{arXiv:1702.08587}  (\bibinfo{year}{2017}).

\bibitem[{\citenamefont{Koschorreck et~al.}(2012)\citenamefont{Koschorreck,
  Pertot, Vogt, Fr\"ohlich, Feld, and K\"ohl}}]{marco12Nature.485.619}
\bibinfo{author}{\bibfnamefont{M.}~\bibnamefont{Koschorreck}},
  \bibinfo{author}{\bibfnamefont{D.}~\bibnamefont{Pertot}},
  \bibinfo{author}{\bibfnamefont{E.}~\bibnamefont{Vogt}},
  \bibinfo{author}{\bibfnamefont{B.}~\bibnamefont{Fr\"ohlich}},
  \bibinfo{author}{\bibfnamefont{M.}~\bibnamefont{Feld}}, \bibnamefont{and}
  \bibinfo{author}{\bibfnamefont{M.}~\bibnamefont{K\"ohl}},
  \bibinfo{journal}{Nature} \textbf{\bibinfo{volume}{485}},
  \bibinfo{pages}{619} (\bibinfo{year}{2012}).

\bibitem[{\citenamefont{Zhang et~al.}(2012)\citenamefont{Zhang, Ong, Arakelyan,
  and Thomas}}]{zhang12PhysRevLett.108.235302}
\bibinfo{author}{\bibfnamefont{Y.}~\bibnamefont{Zhang}},
  \bibinfo{author}{\bibfnamefont{W.}~\bibnamefont{Ong}},
  \bibinfo{author}{\bibfnamefont{I.}~\bibnamefont{Arakelyan}},
  \bibnamefont{and} \bibinfo{author}{\bibfnamefont{J.~E.}
  \bibnamefont{Thomas}}, \bibinfo{journal}{Phys. Rev. Lett.}
  \textbf{\bibinfo{volume}{108}}, \bibinfo{pages}{235302}
  (\bibinfo{year}{2012}),

\bibitem[{\citenamefont{Kohstall et~al.}(2012)\citenamefont{Kohstall, Zaccanti,
  Jag, Trenkwalder, Massignan, Bruun, Schreck, and
  Grimm}}]{kohstall12Nature.485.615}
\bibinfo{author}{\bibfnamefont{C.}~\bibnamefont{Kohstall}},
  \bibinfo{author}{\bibfnamefont{M.}~\bibnamefont{Zaccanti}},
  \bibinfo{author}{\bibfnamefont{M.}~\bibnamefont{Jag}},
  \bibinfo{author}{\bibfnamefont{A.}~\bibnamefont{Trenkwalder}},
  \bibinfo{author}{\bibfnamefont{P.}~\bibnamefont{Massignan}},
  \bibinfo{author}{\bibfnamefont{G.~M.} \bibnamefont{Bruun}},
  \bibinfo{author}{\bibfnamefont{F.}~\bibnamefont{Schreck}}, \bibnamefont{and}
  \bibinfo{author}{\bibfnamefont{R.}~\bibnamefont{Grimm}},
  \bibinfo{journal}{Nature} \textbf{\bibinfo{volume}{485}},
  \bibinfo{pages}{615} (\bibinfo{year}{2012}).

\bibitem[{\citenamefont{Cetina et~al.}(2016)\citenamefont{Cetina, Jag, Lous,
  Fritsche, Walraven, Grimm, Levinsen, Parish, Schmidt, Knap
  et~al.}}]{cetina16Science.354.6308}
\bibinfo{author}{\bibfnamefont{M.}~\bibnamefont{Cetina}},
  \bibinfo{author}{\bibfnamefont{M.}~\bibnamefont{Jag}},
  \bibinfo{author}{\bibfnamefont{R.~S.} \bibnamefont{Lous}},
  \bibinfo{author}{\bibfnamefont{I.}~\bibnamefont{Fritsche}},
  \bibinfo{author}{\bibfnamefont{J.~T.~M.} \bibnamefont{Walraven}},
  \bibinfo{author}{\bibfnamefont{R.}~\bibnamefont{Grimm}},
  \bibinfo{author}{\bibfnamefont{J.}~\bibnamefont{Levinsen}},
  \bibinfo{author}{\bibfnamefont{M.~M.} \bibnamefont{Parish}},
  \bibinfo{author}{\bibfnamefont{R.}~\bibnamefont{Schmidt}},
  \bibinfo{author}{\bibfnamefont{M.}~\bibnamefont{Knap}}, \bibnamefont{et~al.},
  \bibinfo{journal}{Science} \textbf{\bibinfo{volume}{354}},
  \bibinfo{pages}{96} (\bibinfo{year}{2016}). 

\bibitem[{\citenamefont{Scazza et~al.}(2017)\citenamefont{Scazza, Valtolina,
  Massignan, Recati, Amico, Burchianti, Fort, Inguscio, Zaccanti, and
  Roati}}]{scazza17PhysRevLett.118.083602}
\bibinfo{author}{\bibfnamefont{F.}~\bibnamefont{Scazza}},
  \bibinfo{author}{\bibfnamefont{G.}~\bibnamefont{Valtolina}},
  \bibinfo{author}{\bibfnamefont{P.}~\bibnamefont{Massignan}},
  \bibinfo{author}{\bibfnamefont{A.}~\bibnamefont{Recati}},
  \bibinfo{author}{\bibfnamefont{A.}~\bibnamefont{Amico}},
  \bibinfo{author}{\bibfnamefont{A.}~\bibnamefont{Burchianti}},
  \bibinfo{author}{\bibfnamefont{C.}~\bibnamefont{Fort}},
  \bibinfo{author}{\bibfnamefont{M.}~\bibnamefont{Inguscio}},
  \bibinfo{author}{\bibfnamefont{M.}~\bibnamefont{Zaccanti}}, \bibnamefont{and}
  \bibinfo{author}{\bibfnamefont{G.}~\bibnamefont{Roati}},
  \bibinfo{journal}{Phys. Rev. Lett.} \textbf{\bibinfo{volume}{118}},
  \bibinfo{pages}{083602} (\bibinfo{year}{2017}),

\bibitem[{\citenamefont{Casteels et~al.}(2011)\citenamefont{Casteels, Cauteren,
  Tempere, and Devreese}}]{casteels11LaserPhysics.21.1480}
\bibinfo{author}{\bibfnamefont{W.}~\bibnamefont{Casteels}},
  \bibinfo{author}{\bibfnamefont{T.}~\bibnamefont{Cauteren}},
  \bibinfo{author}{\bibfnamefont{J.}~\bibnamefont{Tempere}}, \bibnamefont{and}
  \bibinfo{author}{\bibfnamefont{J.~T.} \bibnamefont{Devreese}},
  \bibinfo{journal}{Laser Physics} \textbf{\bibinfo{volume}{21}},
  \bibinfo{pages}{1480} (\bibinfo{year}{2011}). 

\bibitem[{\citenamefont{Casteels et~al.}(2012)\citenamefont{Casteels, Tempere,
  and Devreese}}]{casteels12PhysRevA.86.043614}
\bibinfo{author}{\bibfnamefont{W.}~\bibnamefont{Casteels}},
  \bibinfo{author}{\bibfnamefont{J.}~\bibnamefont{Tempere}}, \bibnamefont{and}
  \bibinfo{author}{\bibfnamefont{J.~T.} \bibnamefont{Devreese}},
  \bibinfo{journal}{Phys. Rev. A} \textbf{\bibinfo{volume}{86}},
  \bibinfo{pages}{043614} (\bibinfo{year}{2012}).

\bibitem[{\citenamefont{Rath and Schmidt}(2013)}]{rath13PhysRevA.88.053632}
\bibinfo{author}{\bibfnamefont{S.~P.} \bibnamefont{Rath}} \bibnamefont{and}
  \bibinfo{author}{\bibfnamefont{R.}~\bibnamefont{Schmidt}},
  \bibinfo{journal}{Phys. Rev. A} \textbf{\bibinfo{volume}{88}},
  \bibinfo{pages}{053632} (\bibinfo{year}{2013}).

\bibitem[{\citenamefont{Kain and Ling}(2014)}]{kain14PhysRevA.89.023612}
\bibinfo{author}{\bibfnamefont{B.}~\bibnamefont{Kain}} \bibnamefont{and}
  \bibinfo{author}{\bibfnamefont{H.~Y.} \bibnamefont{Ling}},
  \bibinfo{journal}{Phys. Rev. A} \textbf{\bibinfo{volume}{89}},
  \bibinfo{pages}{023612} (\bibinfo{year}{2014}).

\bibitem[{\citenamefont{Shashi et~al.}(2014)\citenamefont{Shashi, Grusdt,
  Abanin, and Demler}}]{shashi14PhysRevA.89.053617}
\bibinfo{author}{\bibfnamefont{A.}~\bibnamefont{Shashi}},
  \bibinfo{author}{\bibfnamefont{F.}~\bibnamefont{Grusdt}},
  \bibinfo{author}{\bibfnamefont{D.~A.} \bibnamefont{Abanin}},
  \bibnamefont{and} \bibinfo{author}{\bibfnamefont{E.}~\bibnamefont{Demler}},
  \bibinfo{journal}{Phys. Rev. A} \textbf{\bibinfo{volume}{89}},
  \bibinfo{pages}{053617} (\bibinfo{year}{2014}).

\bibitem[{\citenamefont{Li and Das~Sarma}(2014)}]{li14PhysRevA.90.013618}
\bibinfo{author}{\bibfnamefont{W.}~\bibnamefont{Li}} \bibnamefont{and}
  \bibinfo{author}{\bibfnamefont{S.}~\bibnamefont{Das~Sarma}},
  \bibinfo{journal}{Phys. Rev. A} \textbf{\bibinfo{volume}{90}},
  \bibinfo{pages}{013618} (\bibinfo{year}{2014}).

\bibitem[{\citenamefont{Grusdt et~al.}(2015)\citenamefont{Grusdt, Shchadilova,
  Rubtsov, and Demler}}]{grusdt15ScientificReports.5.12124}
\bibinfo{author}{\bibfnamefont{F.}~\bibnamefont{Grusdt}},
  \bibinfo{author}{\bibfnamefont{Y.~E.} \bibnamefont{Shchadilova}},
  \bibinfo{author}{\bibfnamefont{A.~N.} \bibnamefont{Rubtsov}},
  \bibnamefont{and} \bibinfo{author}{\bibfnamefont{E.}~\bibnamefont{Demler}},
  \bibinfo{journal}{Scientific Reports} \textbf{\bibinfo{volume}{5}},
  \bibinfo{pages}{12124} (\bibinfo{year}{2015}).

\bibitem[{\citenamefont{Ardila and
  Giorgini}(2015)}]{ardila15PhysRevA.92.033612}
\bibinfo{author}{\bibfnamefont{L.~A. P.~n.} \bibnamefont{Ardila}}
  \bibnamefont{and} \bibinfo{author}{\bibfnamefont{S.}~\bibnamefont{Giorgini}},
  \bibinfo{journal}{Phys. Rev. A} \textbf{\bibinfo{volume}{92}},
  \bibinfo{pages}{033612} (\bibinfo{year}{2015}),

\bibitem[{\citenamefont{Vlietinck et~al.}(2015)\citenamefont{Vlietinck,
  Casteels, Houcke, Tempere, Ryckebusch, and
  Devreese}}]{vlietinck15NewJournalOfPhysics.17.033023}
\bibinfo{author}{\bibfnamefont{J.}~\bibnamefont{Vlietinck}},
  \bibinfo{author}{\bibfnamefont{W.}~\bibnamefont{Casteels}},
  \bibinfo{author}{\bibfnamefont{K.~V.} \bibnamefont{Houcke}},
  \bibinfo{author}{\bibfnamefont{J.}~\bibnamefont{Tempere}},
  \bibinfo{author}{\bibfnamefont{J.}~\bibnamefont{Ryckebusch}},
  \bibnamefont{and} \bibinfo{author}{\bibfnamefont{J.~T.}
  \bibnamefont{Devreese}}, \bibinfo{journal}{New Journal of Physics}
  \textbf{\bibinfo{volume}{17}}, \bibinfo{pages}{033023}
  (\bibinfo{year}{2015}).

\bibitem[{\citenamefont{Shchadilova
  et~al.}(2016{\natexlab{a}})\citenamefont{Shchadilova, Grusdt, Rubtsov, and
  Demler}}]{shchadilova16PhysRevA.93.043606}
\bibinfo{author}{\bibfnamefont{Y.~E.} \bibnamefont{Shchadilova}},
  \bibinfo{author}{\bibfnamefont{F.}~\bibnamefont{Grusdt}},
  \bibinfo{author}{\bibfnamefont{A.~N.} \bibnamefont{Rubtsov}},
  \bibnamefont{and} \bibinfo{author}{\bibfnamefont{E.}~\bibnamefont{Demler}},
  \bibinfo{journal}{Phys. Rev. A} \textbf{\bibinfo{volume}{93}},
  \bibinfo{pages}{043606} (\bibinfo{year}{2016}{\natexlab{a}}),

\bibitem[{\citenamefont{Christensen et~al.}(2015)\citenamefont{Christensen,
  Levinsen, and Bruun}}]{sogaard15PhysRevLett.115.160401}
\bibinfo{author}{\bibfnamefont{R.~S.} \bibnamefont{Christensen}},
  \bibinfo{author}{\bibfnamefont{J.}~\bibnamefont{Levinsen}}, \bibnamefont{and}
  \bibinfo{author}{\bibfnamefont{G.~M.} \bibnamefont{Bruun}},
  \bibinfo{journal}{Phys. Rev. Lett.} \textbf{\bibinfo{volume}{115}},
  \bibinfo{pages}{160401} (\bibinfo{year}{2015}).

\bibitem[{\citenamefont{Levinsen et~al.}(2015)\citenamefont{Levinsen, Parish,
  and Bruun}}]{levinsen15PhysRevLett.115.125302}
\bibinfo{author}{\bibfnamefont{J.}~\bibnamefont{Levinsen}},
  \bibinfo{author}{\bibfnamefont{M.~M.} \bibnamefont{Parish}},
  \bibnamefont{and} \bibinfo{author}{\bibfnamefont{G.~M.} \bibnamefont{Bruun}},
  \bibinfo{journal}{Phys. Rev. Lett.} \textbf{\bibinfo{volume}{115}},
  \bibinfo{pages}{125302} (\bibinfo{year}{2015}).

\bibitem[{\citenamefont{Kain and Ling}(2016)}]{kain16PhysRevA.94.013621}
\bibinfo{author}{\bibfnamefont{B.}~\bibnamefont{Kain}} \bibnamefont{and}
  \bibinfo{author}{\bibfnamefont{H.~Y.} \bibnamefont{Ling}},
  \bibinfo{journal}{Phys. Rev. A} \textbf{\bibinfo{volume}{94}},
  \bibinfo{pages}{013621} (\bibinfo{year}{2016}),

\bibitem[{\citenamefont{Schmidt et~al.}(2016)\citenamefont{Schmidt, Sadeghpour,
  and Demler}}]{schmidt16PhysRevLett.116.105302}
\bibinfo{author}{\bibfnamefont{R.}~\bibnamefont{Schmidt}},
  \bibinfo{author}{\bibfnamefont{H.~R.} \bibnamefont{Sadeghpour}},
  \bibnamefont{and} \bibinfo{author}{\bibfnamefont{E.}~\bibnamefont{Demler}},
  \bibinfo{journal}{Phys. Rev. Lett.} \textbf{\bibinfo{volume}{116}},
  \bibinfo{pages}{105302} (\bibinfo{year}{2016}),

\bibitem[{\citenamefont{Shchadilova
  et~al.}(2016{\natexlab{b}})\citenamefont{Shchadilova, Schmidt, Grusdt, and
  Demler}}]{shchadilova16PhysRevLett.117.113002}
\bibinfo{author}{\bibfnamefont{Y.~E.} \bibnamefont{Shchadilova}},
  \bibinfo{author}{\bibfnamefont{R.}~\bibnamefont{Schmidt}},
  \bibinfo{author}{\bibfnamefont{F.}~\bibnamefont{Grusdt}}, \bibnamefont{and}
  \bibinfo{author}{\bibfnamefont{E.}~\bibnamefont{Demler}},
  \bibinfo{journal}{Phys. Rev. Lett.} \textbf{\bibinfo{volume}{117}},
  \bibinfo{pages}{113002} (\bibinfo{year}{2016}{\natexlab{b}}),
  
\bibitem{Grusdt17arXiv:1704.02605}
	F.\ Grusdt, R.\ Schmidt, Y.\ E.\ Shchadilova, and E.\ A.\ Demler, Phys.\ Rev.\ A \textbf{96}, 013607 (2017).

\bibitem[{\citenamefont{Grusdt et~al.}(2017{\natexlab{b}})\citenamefont{Grusdt,
  Astrakharchik, and Demler}}]{Grusdt17arXiv:1704.02606}
\bibinfo{author}{\bibfnamefont{F.}~\bibnamefont{Grusdt}},
  \bibinfo{author}{\bibfnamefont{G.~E.} \bibnamefont{Astrakharchik}},
  \bibnamefont{and} \bibinfo{author}{\bibfnamefont{E.~A.}
  \bibnamefont{Demler}}, \bibinfo{journal}{arXiv:1704.02606}
  (\bibinfo{year}{2017}{\natexlab{b}}).
  

\bibitem{Sun}
	M.\ Sun, H.\ Zhai, and X.\ Cui, Phys.\ Rev.\ Lett.\ \textbf{119}, 013401 (2017).
	
\bibitem{Sun2}
	M.\ Sun and X.\ Cui, Phys.\ Rev.\ A \textbf{96}, 022707 (2017).

\bibitem[{\citenamefont{Catani et~al.}(2012)\citenamefont{Catani, Lamporesi,
  Naik, Gring, Inguscio, Minardi, Kantian, and
  Giamarchi}}]{catani12PhysRevA.85.023623}
\bibinfo{author}{\bibfnamefont{J.}~\bibnamefont{Catani}},
  \bibinfo{author}{\bibfnamefont{G.}~\bibnamefont{Lamporesi}},
  \bibinfo{author}{\bibfnamefont{D.}~\bibnamefont{Naik}},
  \bibinfo{author}{\bibfnamefont{M.}~\bibnamefont{Gring}},
  \bibinfo{author}{\bibfnamefont{M.}~\bibnamefont{Inguscio}},
  \bibinfo{author}{\bibfnamefont{F.}~\bibnamefont{Minardi}},
  \bibinfo{author}{\bibfnamefont{A.}~\bibnamefont{Kantian}}, \bibnamefont{and}
  \bibinfo{author}{\bibfnamefont{T.}~\bibnamefont{Giamarchi}},
  \bibinfo{journal}{Phys. Rev. A} \textbf{\bibinfo{volume}{85}},
  \bibinfo{pages}{023623} (\bibinfo{year}{2012}).

\bibitem[{\citenamefont{Hu et~al.}(2016)\citenamefont{Hu, Van~de Graaff, Kedar,
  Corson, Cornell, and Jin}}]{hu16PhysRevLett.117.055301}
\bibinfo{author}{\bibfnamefont{M.-G.} \bibnamefont{Hu}},
  \bibinfo{author}{\bibfnamefont{M.~J.} \bibnamefont{Van~de Graaff}},
  \bibinfo{author}{\bibfnamefont{D.}~\bibnamefont{Kedar}},
  \bibinfo{author}{\bibfnamefont{J.~P.} \bibnamefont{Corson}},
  \bibinfo{author}{\bibfnamefont{E.~A.} \bibnamefont{Cornell}},
  \bibnamefont{and} \bibinfo{author}{\bibfnamefont{D.~S.} \bibnamefont{Jin}},
  \bibinfo{journal}{Phys. Rev. Lett.} \textbf{\bibinfo{volume}{117}},
  \bibinfo{pages}{055301} (\bibinfo{year}{2016}),

\bibitem[{\citenamefont{J\o{}rgensen et~al.}(2016)\citenamefont{J\o{}rgensen,
  Wacker, Skalmstang, Parish, Levinsen, Christensen, Bruun, and
  Arlt}}]{jorgensen16PhysRevLett.117.055302}
\bibinfo{author}{\bibfnamefont{N.~B.} \bibnamefont{J\o{}rgensen}},
  \bibinfo{author}{\bibfnamefont{L.}~\bibnamefont{Wacker}},
  \bibinfo{author}{\bibfnamefont{K.~T.} \bibnamefont{Skalmstang}},
  \bibinfo{author}{\bibfnamefont{M.~M.} \bibnamefont{Parish}},
  \bibinfo{author}{\bibfnamefont{J.}~\bibnamefont{Levinsen}},
  \bibinfo{author}{\bibfnamefont{R.~S.} \bibnamefont{Christensen}},
  \bibinfo{author}{\bibfnamefont{G.~M.} \bibnamefont{Bruun}}, \bibnamefont{and}
  \bibinfo{author}{\bibfnamefont{J.~J.} \bibnamefont{Arlt}},
  \bibinfo{journal}{Phys. Rev. Lett.} \textbf{\bibinfo{volume}{117}},
  \bibinfo{pages}{055302} (\bibinfo{year}{2016}),

\bibitem[{\citenamefont{Lee et~al.}(1953)\citenamefont{Lee, Low, and
  Pines}}]{lee53PhysRev.90.297}
\bibinfo{author}{\bibfnamefont{T.~D.} \bibnamefont{Lee}},
  \bibinfo{author}{\bibfnamefont{F.~E.} \bibnamefont{Low}}, \bibnamefont{and}
  \bibinfo{author}{\bibfnamefont{D.}~\bibnamefont{Pines}},
  \bibinfo{journal}{Phys. Rev.} \textbf{\bibinfo{volume}{90}},
  \bibinfo{pages}{297} (\bibinfo{year}{1953}).

\bibitem[{\citenamefont{Edwards}(1990)}]{Edwards1990PThPS.101.453}
\bibinfo{author}{\bibfnamefont{D.~M.} \bibnamefont{Edwards}},
  \bibinfo{journal}{Progress of Theoretical Physics Supplement}
  \textbf{\bibinfo{volume}{101}}, \bibinfo{pages}{453} (\bibinfo{year}{1990}).

\bibitem[{\citenamefont{Castella and
  Zotos}(1993)}]{Castella93PhysRevB.47.16186}
\bibinfo{author}{\bibfnamefont{H.}~\bibnamefont{Castella}} \bibnamefont{and}
  \bibinfo{author}{\bibfnamefont{X.}~\bibnamefont{Zotos}},
  \bibinfo{journal}{Phys. Rev. B} \textbf{\bibinfo{volume}{47}},
  \bibinfo{pages}{16186} (\bibinfo{year}{1993}),

\bibitem[{\citenamefont{Lamacraft}(2009)}]{Lamacraft09PhysRevB.79.241105}
\bibinfo{author}{\bibfnamefont{A.}~\bibnamefont{Lamacraft}},
  \bibinfo{journal}{Phys. Rev. B} \textbf{\bibinfo{volume}{79}},
  \bibinfo{pages}{241105} (\bibinfo{year}{2009}),

\bibitem[{\citenamefont{Mathy et~al.}(2012)\citenamefont{Mathy, Zvonarev, and
  Demler}}]{Mathy12NaturePhysics.8.881}
\bibinfo{author}{\bibfnamefont{C.~J.~M.} \bibnamefont{Mathy}},
  \bibinfo{author}{\bibfnamefont{M.~B.} \bibnamefont{Zvonarev}},
  \bibnamefont{and} \bibinfo{author}{\bibfnamefont{E.}~\bibnamefont{Demler}},
  \bibinfo{journal}{Nature Physics} \textbf{\bibinfo{volume}{8}},
  \bibinfo{pages}{881} (\bibinfo{year}{2012}),

\bibitem[{\citenamefont{Edwards}(2013)}]{edward13JournalOfPhysics.25.425602}
\bibinfo{author}{\bibfnamefont{D.~M.} \bibnamefont{Edwards}},
  \bibinfo{journal}{Journal of Physics: Condensed Matter}
  \textbf{\bibinfo{volume}{25}}, \bibinfo{pages}{425602}
  (\bibinfo{year}{2013}),

\bibitem[{\citenamefont{Grusdt et~al.}(2016)\citenamefont{Grusdt, Yao, Abanin,
  Fleischhauer, and Demler}}]{grusdt17NatCommun.7.11994}
\bibinfo{author}{\bibfnamefont{F.}~\bibnamefont{Grusdt}},
  \bibinfo{author}{\bibfnamefont{N.~Y.} \bibnamefont{Yao}},
  \bibinfo{author}{\bibfnamefont{D.}~\bibnamefont{Abanin}},
  \bibinfo{author}{\bibfnamefont{M.}~\bibnamefont{Fleischhauer}},
  \bibnamefont{and} \bibinfo{author}{\bibfnamefont{E.}~\bibnamefont{Demler}},
  \bibinfo{journal}{Nat. Commun.} \textbf{\bibinfo{volume}{7}},
  \bibinfo{pages}{11994} (\bibinfo{year}{2016}),

\bibitem[{\citenamefont{McGuire}(1965)}]{McGuire65JMathPhys.6.432}
\bibinfo{author}{\bibfnamefont{J.~B.} \bibnamefont{McGuire}},
  \bibinfo{journal}{J. Math. Phys.} \textbf{\bibinfo{volume}{6}},
  \bibinfo{pages}{432} (\bibinfo{year}{1965}),

\bibitem[{\citenamefont{McGuire}(1966)}]{McGuire66JMathPhys.7.123}
\bibinfo{author}{\bibfnamefont{J.~B.} \bibnamefont{McGuire}},
  \bibinfo{journal}{J. Math. Phys.} \textbf{\bibinfo{volume}{7}},
  \bibinfo{pages}{123} (\bibinfo{year}{1966}),

\bibitem[{\citenamefont{Wigner and Seitz}(1934)}]{wigner34PhysRev.46.509}
\bibinfo{author}{\bibfnamefont{E.}~\bibnamefont{Wigner}} \bibnamefont{and}
  \bibinfo{author}{\bibfnamefont{F.}~\bibnamefont{Seitz}},
  \bibinfo{journal}{Phys. Rev.} \textbf{\bibinfo{volume}{46}},
  \bibinfo{pages}{509} (\bibinfo{year}{1934}),

\bibitem{Chen Guan}
	L.\ Guan and S.\ Chen, Phys.\ Rev.\ Lett.\ \textbf{105}, 175301 (2010).

\bibitem[{\citenamefont{Olshanii}(1998)}]{olshanii98PhysRevLett.81.938}
\bibinfo{author}{\bibfnamefont{M.}~\bibnamefont{Olshanii}},
  \bibinfo{journal}{Phys. Rev. Lett.} \textbf{\bibinfo{volume}{81}},
  \bibinfo{pages}{938} (\bibinfo{year}{1998}).

\bibitem[{\citenamefont{Bergeman et~al.}(2003)\citenamefont{Bergeman, Moore,
  and Olshanii}}]{bergeman03PhysRevLett.91.163201}
\bibinfo{author}{\bibfnamefont{T.}~\bibnamefont{Bergeman}},
  \bibinfo{author}{\bibfnamefont{M.~G.} \bibnamefont{Moore}}, \bibnamefont{and}
  \bibinfo{author}{\bibfnamefont{M.}~\bibnamefont{Olshanii}},
  \bibinfo{journal}{Phys. Rev. Lett.} \textbf{\bibinfo{volume}{91}},
  \bibinfo{pages}{163201} (\bibinfo{year}{2003}).

\bibitem[{\citenamefont{Peano et~al.}(2005)\citenamefont{Peano, Thorwart, Mora,
  and Egger}}]{peano05NewJournalOfPhysics.7.192}
\bibinfo{author}{\bibfnamefont{V.}~\bibnamefont{Peano}},
  \bibinfo{author}{\bibfnamefont{M.}~\bibnamefont{Thorwart}},
  \bibinfo{author}{\bibfnamefont{C.}~\bibnamefont{Mora}}, \bibnamefont{and}
  \bibinfo{author}{\bibfnamefont{R.}~\bibnamefont{Egger}},
  \bibinfo{journal}{New Journal of Physics} \textbf{\bibinfo{volume}{7}},
  \bibinfo{pages}{192} (\bibinfo{year}{2005}),

\bibitem[{\citenamefont{Guan et~al.}(2013)\citenamefont{Guan, Batchelor, and
  Lee}}]{Guan13RevModPhys.85.1633}
\bibinfo{author}{\bibfnamefont{X.-W.} \bibnamefont{Guan}},
  \bibinfo{author}{\bibfnamefont{M.~T.} \bibnamefont{Batchelor}},
  \bibnamefont{and} \bibinfo{author}{\bibfnamefont{C.}~\bibnamefont{Lee}},
  \bibinfo{journal}{Rev. Mod. Phys.} \textbf{\bibinfo{volume}{85}},
  \bibinfo{pages}{1633} (\bibinfo{year}{2013}),

\bibitem[{\citenamefont{Giraud and
  Combescot}(2009)}]{giraud09PhysRevA.79.043615}
\bibinfo{author}{\bibfnamefont{S.}~\bibnamefont{Giraud}} \bibnamefont{and}
  \bibinfo{author}{\bibfnamefont{R.}~\bibnamefont{Combescot}},
  \bibinfo{journal}{Phys. Rev. A} \textbf{\bibinfo{volume}{79}},
  \bibinfo{pages}{043615} (\bibinfo{year}{2009}),

\bibitem[{\citenamefont{Bethe}(1931)}]{Bethe1931Zphys.71.205}
\bibinfo{author}{\bibfnamefont{H.~A.} \bibnamefont{Bethe}},
  \bibinfo{journal}{Z. Phys.} \textbf{\bibinfo{volume}{71}},
  \bibinfo{pages}{205} (\bibinfo{year}{1931}).

\bibitem[{\citenamefont{Yang}(1967)}]{yang67PhysRevLett.19.1312}
\bibinfo{author}{\bibfnamefont{C.~N.} \bibnamefont{Yang}},
  \bibinfo{journal}{Phys. Rev. Lett.} \textbf{\bibinfo{volume}{19}},
  \bibinfo{pages}{1312} (\bibinfo{year}{1967}),

\bibitem[{\citenamefont{Gaudin}(1967)}]{gaudin67PhysLettA.24.55}
\bibinfo{author}{\bibfnamefont{M.}~\bibnamefont{Gaudin}},
  \bibinfo{journal}{Phys. Lett. A} \textbf{\bibinfo{volume}{24}},
  \bibinfo{pages}{55} (\bibinfo{year}{1967}).

\bibitem[{\citenamefont{Lieb and Wu}(1968)}]{lieb68PhysRevLett.20.1445}
\bibinfo{author}{\bibfnamefont{E.~H.} \bibnamefont{Lieb}} \bibnamefont{and}
  \bibinfo{author}{\bibfnamefont{F.~Y.} \bibnamefont{Wu}},
  \bibinfo{journal}{Phys. Rev. Lett.} \textbf{\bibinfo{volume}{20}},
  \bibinfo{pages}{1445} (\bibinfo{year}{1968}),

\bibitem[{\citenamefont{Altman et~al.}(2004)\citenamefont{Altman, Demler, and
  Lukin}}]{ehud04PhysRevA.70.013603}
\bibinfo{author}{\bibfnamefont{E.}~\bibnamefont{Altman}},
  \bibinfo{author}{\bibfnamefont{E.}~\bibnamefont{Demler}}, \bibnamefont{and}
  \bibinfo{author}{\bibfnamefont{M.~D.} \bibnamefont{Lukin}},
  \bibinfo{journal}{Phys. Rev. A} \textbf{\bibinfo{volume}{70}},
  \bibinfo{pages}{013603} (\bibinfo{year}{2004}).

\bibitem[{\citenamefont{Folling et~al.}(2005)\citenamefont{Folling, Gerbier,
  Widera, Mandel, Gericke, and Bloch}}]{simon05Nature.434.481}
\bibinfo{author}{\bibfnamefont{S.}~\bibnamefont{Folling}},
  \bibinfo{author}{\bibfnamefont{F.}~\bibnamefont{Gerbier}},
  \bibinfo{author}{\bibfnamefont{A.}~\bibnamefont{Widera}},
  \bibinfo{author}{\bibfnamefont{O.}~\bibnamefont{Mandel}},
  \bibinfo{author}{\bibfnamefont{T.}~\bibnamefont{Gericke}}, \bibnamefont{and}
  \bibinfo{author}{\bibfnamefont{I.}~\bibnamefont{Bloch}},
  \bibinfo{journal}{Nature} \textbf{\bibinfo{volume}{434}},
  \bibinfo{pages}{481} (\bibinfo{year}{2005}).

\bibitem[{\citenamefont{Rom et~al.}(2006)\citenamefont{Rom, Best, van Oosten,
  Schneider, Folling, Paredes, and Bloch}}]{rom06Nature.434.481}
\bibinfo{author}{\bibfnamefont{T.}~\bibnamefont{Rom}},
  \bibinfo{author}{\bibfnamefont{T.}~\bibnamefont{Best}},
  \bibinfo{author}{\bibfnamefont{D.}~\bibnamefont{van Oosten}},
  \bibinfo{author}{\bibfnamefont{U.}~\bibnamefont{Schneider}},
  \bibinfo{author}{\bibfnamefont{S.}~\bibnamefont{Folling}},
  \bibinfo{author}{\bibfnamefont{B.}~\bibnamefont{Paredes}}, \bibnamefont{and}
  \bibinfo{author}{\bibfnamefont{I.}~\bibnamefont{Bloch}},
  \bibinfo{journal}{Nature} \textbf{\bibinfo{volume}{444}},
  \bibinfo{pages}{733} (\bibinfo{year}{2006}).

\bibitem[{\citenamefont{Chin et~al.}(2004)\citenamefont{Chin, Bartenstein,
  Altmeyer, Riedl, Jochim, Denschlag, and Grimm}}]{chin04Science305.20082004}
\bibinfo{author}{\bibfnamefont{C.}~\bibnamefont{Chin}},
  \bibinfo{author}{\bibfnamefont{M.}~\bibnamefont{Bartenstein}},
  \bibinfo{author}{\bibfnamefont{A.}~\bibnamefont{Altmeyer}},
  \bibinfo{author}{\bibfnamefont{S.}~\bibnamefont{Riedl}},
  \bibinfo{author}{\bibfnamefont{S.}~\bibnamefont{Jochim}},
  \bibinfo{author}{\bibfnamefont{J.~H.} \bibnamefont{Denschlag}},
  \bibnamefont{and} \bibinfo{author}{\bibfnamefont{R.}~\bibnamefont{Grimm}},
  \bibinfo{journal}{Science} \textbf{\bibinfo{volume}{305}},
  \bibinfo{pages}{1128} (\bibinfo{year}{2004}).

\bibitem[{\citenamefont{Kinnunen et~al.}(2004)\citenamefont{Kinnunen,
  Rodriguez, and T{\"o}rm{\"a}}}]{kinnunen04Science305.20082004}
\bibinfo{author}{\bibfnamefont{J.}~\bibnamefont{Kinnunen}},
  \bibinfo{author}{\bibfnamefont{M.}~\bibnamefont{Rodriguez}},
  \bibnamefont{and}
  \bibinfo{author}{\bibfnamefont{P.}~\bibnamefont{T{\"o}rm{\"a}}},
  \bibinfo{journal}{Science} \textbf{\bibinfo{volume}{305}},
  \bibinfo{pages}{1131} (\bibinfo{year}{2004}).

\bibitem[{\citenamefont{Grusdt and Demler}(2015)}]{grusdt15arXiv:1510.04934}
\bibinfo{author}{\bibfnamefont{F.}~\bibnamefont{Grusdt}} \bibnamefont{and}
  \bibinfo{author}{\bibfnamefont{E.}~\bibnamefont{Demler}},
  \bibinfo{journal}{arXiv:1510.04934}  (\bibinfo{year}{2015}).

\end{thebibliography}

\end{document}